\documentclass[pra,aps,twocolumn,floatfix,longbibliography]{revtex4-2}

\usepackage{amsmath}
\usepackage{amssymb}
\usepackage{txfonts}
\usepackage{microtype}
\usepackage{graphicx}
\usepackage{color}
\usepackage{ulem}
\usepackage{bm}

\usepackage{diagbox}

\newcommand{\comm}[1]{}

\begin{document}

\title{{ Quasiperiodic nondipole ionization dynamics in the x-ray stabilization regime}}

\author{Aleksandr V. Boitsov}
\email{  boitsov@mpi-hd.mpg.de}
\author{Karen Z. Hatsagortsyan}
\email{ k.hatsagortsyan@mpi-hd.mpg.de}
 \author{Christoph~H.~Keitel}
\affiliation{Max-Planck-Institut f\"{u}r Kernphysik, Saupfercheckweg 1, 69117 Heidelberg, Germany}

\date{\today}

\begin{abstract}

Recent advances in strong x-ray laser techniques enable the study of nonlinear multiphoton ionization in extreme high-frequency fields. Although the stabilization regime in such fields is theoretically established, its modified properties in the nondipole regime for long laser pulses remains unknown. Here, we numerically investigate the strong-field ionization of an atom in a long XUV laser pulse in the nondipole regime. Our study of the time-dependent quantum dynamics reveals a  quasiperiodic modulation of the ionization yield as a function of pulse duration. We demonstrate that the Coulomb-field-induced slow oscillation of the ionized electron wave packet during the interaction is responsible for the observed  modulation of the ionization yield. Furthermore, we scrutinize the unusual photon momentum sharing between the photoelectron and the ion in this extreme regime. These  effects are observable in upcoming x-ray free-electron laser facilities.

\end{abstract}

\date{\today}
\maketitle

\section{Introduction}

Extremely nonlinear phenomena of the strong field  ionization are well investigated with infrared and mid-infrared laser fields resulting in the emergence of attoscience \cite{Becker_2002,Agostini_2004,Corkum_2007,Krausz_2009,Krausz_2024,L'Huillier_2024,Agostini_2024}.
However, recent advancement of strong laser techniques in the XUV and x-ray domain, in particular, the development of XFEL facilities  in DESY \cite{Altarelli_2015}, SLAC \cite{Bostedt_2016}, and other places over the world \cite{Yabashi_2013,Fan_2022}, has raised hopes to achieve the nonlinear multiphoton interaction regime of ionization with x-rays \cite{Walker_2024} .

In high-frequency laser fields when  the laser frequency $\omega$ exceeds the atomic ionization energy $I_p$, the strong field ionization  can enter into the so-called stabilization regime \cite{Reed_1990,Patel1999,Gavrila_2002}. In this case, as the electron oscillation amplitude in the laser field $\alpha_0=E_0/\omega^2$ overtakes the atomic size $a_s$, the average effect of the atomic potential on the electron can be described by a dichotomic Kramers-Henneberger potential \cite{Kramers,Henneberger_1968}. Here, $E_0$ is the  laser field amplitude, and atomic units are used throughout. In this regime, the ionization probability saturates, remaining stable despite the increasing laser intensity.  The photoelectron spectrum in the stabilization regime is characterized by a large near-zero energy peak \cite{Azizi_2025} which is followed by the { weak above-threshold ionization (ATI) peaks} \cite{Telnov_Chu_2021}.

The early works already showed that the nondipole effects tend to significantly modify and even suppress the stabilization \cite{Protopapas_1996,Gaier_2002,Staudt_2003,Protopapas_1997, Keitel_1995}. The nondipole effects in a strong laser field   enter into play  when the classical strong field parameter is large { with $a_0 = E_0/(c\omega)\gtrsim   1$} \cite{RMP_2012}, where $c$ is the speed of light. This condition   is easier to achieve at low frequencies, for instance, $a_0\sim 1$ is achieved in infrared laser fields with a wavelength $\lambda=1000$ nm at a laser intensity $I\sim 10^{18}$ W/cm$^2$, while for x-rays $\lambda=10$ nm, this would require $I\sim 10^{22}$ W/cm$^2$, which could be envisaged with the development of x-ray focusing technique  \cite{LCLS,flash,Dziarzhytski_2016,Vassholz_2021,Tong_2022}.

\begin{figure*}
       \centering\includegraphics[width=0.9\textwidth]{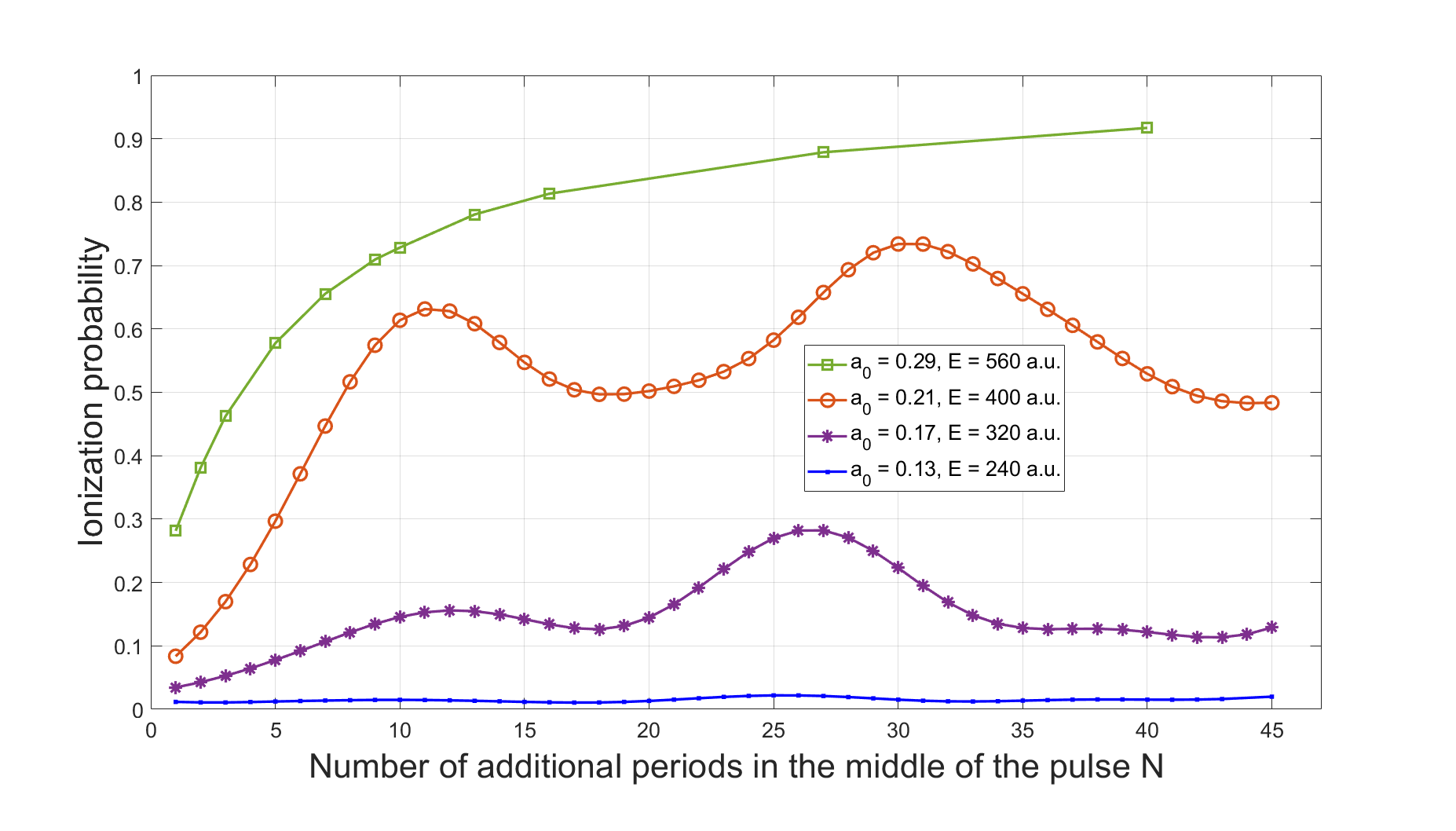}
    \caption{ \textbf{The dependence of the ionization probability of hydrogen-like helium on the laser pulse duration.}  Each line represents a different $a_0$ parameter indicated in the inset, $a_0 = 0.13-0.29$ corresponding to the intensity range of $I=2\times 10^{21}-1.1\times 10^{22} \text{W/cm}^2$. 
    The laser frequency is  $\omega  = 14 \text{ a.u.}$    }
    \label{Fig-1}
\end{figure*}

Ionization in the nondipole regime shows distinctive features already at $a_0 \gtrsim 0.1$ due to the relativistic drift induced by the $v\times B $ Lorentz force. It breaks the symmetry of the system with respect to the laser field direction \cite{Protopapas_1997}, and leaves signatures in the photoelectron momentum distribution (PMD).  The numerical calculations of the time-dependent Schr\"odinger equation (TDSE) in a perturbatively weak x-ray field \cite{Dondera_2012} reveal that the PMD is shifted along the laser propagation direction, similar to ATI in infrared fields \cite{Klaiber_2005,Klaiber_2006}, when the forward ionization is preferred with respect to the backward one. In stronger nonperturbative x-ray fields an interesting interplay between the nondipole and Coulomb fields has been observed in Ref.~\cite{Foerre_2006}, demonstrating counterintuitive features in the photoelectron angular distribution (PAD). In particular, two lobes in PAD, which in the dipole case are perpendicular to the laser wave propagation direction, are bent towards the direction  opposite to the laser propagation, and additionally a new third lobe appears along that direction. This counterintuitive behavior has been explained by the effect  of the Coulomb forces of the atomic core. While the laser magnetic field induces the continuum wave packet drift in the laser propagation direction, the created asymmetry produces a net Coulomb force acting on the wave packet  in the opposite direction, which yields the generation of the anomalous lobe in the PAD.  Later results of Refs.~\cite{Zhou2013,Telnov_Chu_2021} confirmed this effect and noted that the anomalous lobe increases with increasing the laser pulse duration  within limited range. It is worth to underline that the above-mentioned counterintuitive behavior is  general for strong field interaction and has been observed also at infrared laser fields \cite{Ludwig_2014}, when a counterintuitive shift of the peak of the photoelectron momentum distribution (PMD) is observed due to nondipole asymmetry.

The role of the  relativistic effects has been discussed  in Refs.~\cite{Telnov_Chu_2021,Ivanov2017} solving the time-dependent Dirac equation (TDDE). In particular, a spin-flip effect is analyzed in \cite{Ivanov2017} showing that it is extremely small by a factor of $10^{-6}$ at the x-ray field  of $a_0 \approx 0.05-0.15$.
The findings of \cite{Foerre_2006,Zhou2013} have been confirmed in the fully relativistic framework.
Despite a great attention to the nondipole ionization in a strong x-ray field,  the dependence of the ionization probability on the pulse duration largely remained unexplored.

Strong-field ionization of an atom takes place due to absorption of laser photons. In the nondipole regime the photon momentum cannot be neglected. Then a question arises how the linear momentum of absorbed photons is partitioned between the constituents of the atom, between the ionized electron and the emerged ion. While this question is well investigated in the case of the strong-field ionization with infrared laser fields, or in the case of the photoionization \cite{Sommerfeld_1930,Seaton_1995,Smeenk_2011,Klaiber_2013,Chelkowski_2014,Cricchio_2015,Chelkowski_2015,
Hartung_2019,He_2022,Mao_2025} -- single photon ionization with an XUV photon, in the case of the stabilization regime in strong XUV fields, { when the large momentum exchange between the photoelectron and ion via Coulomb interaction non-trivially modifies the momentum sharing,}  has not been addressed.

In this paper, we investigate the role of the laser pulse duration in the strong x-ray field ionization of an hydrogen-like ion in the stabilization regime. By changing the driving pulse duration in a wide range up to 45 optical cycles, we reveal a slow periodic variation of the ionization yield with respect to the pulse duration.  It turns out that the  yield oscillation exists both in the dipole and nondipole regimes, however,  with different oscillation features and with different underlying mechanisms. While in the dipole regime  the oscillation of the ionization yield can be explained by the the internal dynamics of the Kramers-Henneberger (KH) atom along the laser polarization direction \cite{Aynul_2025}, complimented by the dynamic interference phenomenon \cite{Toyota_2007,Toyota_2008, Demekhin_2012,Jiang_2018,Wang_2018, Geng_2021}, in the nondipole regime the dynamics in the laser propagation direction becomes decisive.
The distinctive mechanism, which  brings about the ionization yield oscillation in the nondipole regime, is the slow orbiting of the continuum electron wave packet during the interaction mostly in the propagation direction, which is induced by the nondipole drift force combined with the Coulomb field of the atomic core.
We investigate also the photon momentum partition between the photoelectron and ion in this extreme stabilization regime.  Our investigation is based on the numerical  solution of the  Foldy-Wouthuysen (FW) transformed TDDE in the 2D case in the quasiclassical representation of Silenko \cite{Silenko_2008,Silenko_2013,Silenko_2015}. The numerical calculations are facilitated by the application of the coordinate scaling method \cite{Boitsov_2025}.

The article is structured as follows: in Sec.~\ref{sec:2Dintensities}  the results of our numerical simulations are presented.  The explanation of the yield oscillation in the dipole regime is discussed in Sec.~\ref{sec:dipole_regime}, and in the nondipole regime in Sec.~\ref{sec:nondip}. {   Pulse shape effects  are considered in Sec.~\ref{sec:pulse}.} In Sec.~\ref{sec:momentumSharing} the problem of the photoelectron-ion momentum sharing is addressed. The brief presentation of the calculation method is given in Appendix.

\begin{figure*}
    \centering\includegraphics[width=0.403\textwidth]{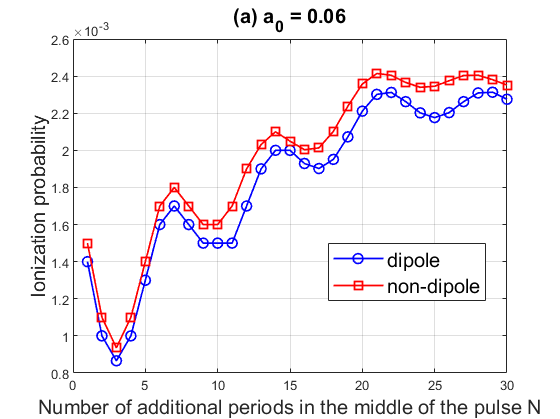}
    \centering\includegraphics[width=0.403\textwidth]{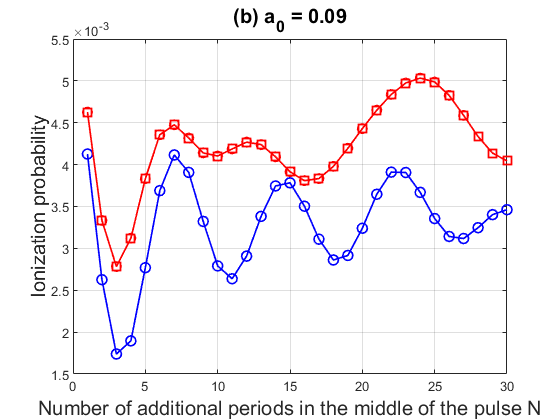}
    \centering\includegraphics[width=0.403\textwidth]{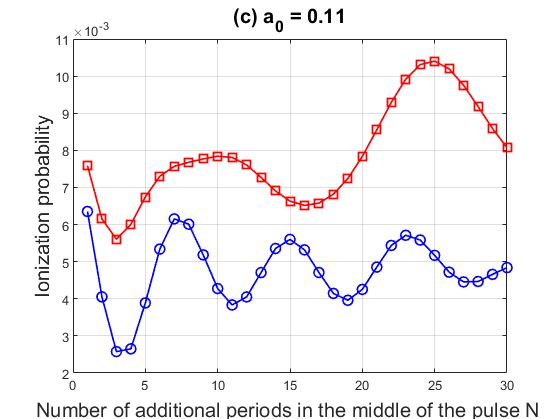}
    \centering\includegraphics[width=0.403\textwidth]{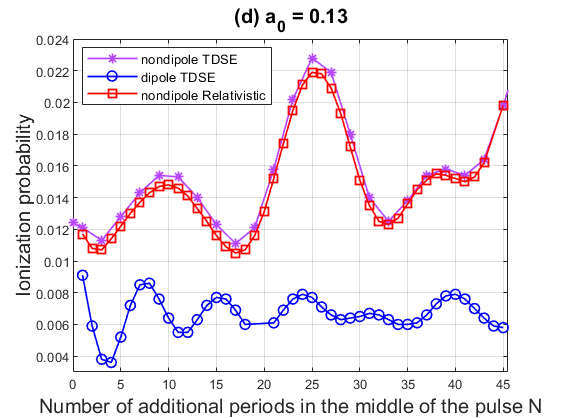}
    \caption{ \textbf{Comparison of the dipole and nondipole calculations for the  ionization probability vs the laser pulse duration.}  (a) $a_0 = 0.06, ~ E_0 = 120 \text{ a.u.}$,   (b) $a_0 = 0.09, ~ E_0 = 180 \text{ a.u.}$, (c) $a_0 = 0.11, ~ E_0 = 210 \text{ a.u.}$, (d) $a_0 = 0.13, ~ E_0 = 240 \text{ a.u.}$; (blue line) -- dipole approximation, (red line) -- nondipole calculations. 
    $\omega  = 14$ a.u.}
    \label{Fig-2}
   \end{figure*}

\section{Ionization yield vs interaction time}\label{sec:2Dintensities}

{We consider the hydrogen-like  atom interaction with a strong laser field in the nondipole stabilization regime, when $\omega>I_p$, $\alpha_0\gtrsim a_s= 1/Z$, with the charge $Z $ of the atomic core.} The following  typical parameters for the stabilization regime  are used:
$\omega  = 14 \text{ a.u.}~(381~\text{eV})$, $ E_0 = 240 - 560 \text{ a.u.}~(I=2\times 10^{21}-1.1\times 10^{22} ~ \text{W/cm}^2)$. In this case we have the strong field parameter $a_0 = 0.13-0.29$, and {the electron oscillation  amplitude $\alpha_0\approx 1.22-2.85$}.
We calculated the ionization yield in laser pulses of different duration  via evaluating the bound state populations after the interaction. Hydrogen-like helium ($Z = 2$) is chosen as a model to have a pronounced Coulomb effect, which we will see is important for the considered effect. {  It is known \cite{Madsen_1999} that in the nonrelativistic limit a scaling law with respect to atomic charge $Z$  exists for the electron dynamics in a Coulomb and laser fields: the dynamics is similar  when appropriately scaling the length $r\rightarrow r/Z$, time $t\rightarrow t/Z^2$,  laser frequency $\omega\rightarrow \omega Z^2$, and the field  $E_0\rightarrow E_0 Z^3$. As the nondipole effects are scaled with $a_0$, and in the considered cases $a_0\lesssim 0.29$ is not large, the same scaling for the process similarity approximately works also in the nondipole case.}

The ionization yield dependence on the laser pulse duration at different laser fields is shown in Fig.~\ref{Fig-1}. The laser pulse is of a trapezoidal form, see Appendix. {  A smooth switching-on and -off is applied in the pulse, with FWHM of $\tau =  2.5 T_0$ (the total switching-on and -off duration is $3\tau$), and the duration of the flat part of the pulse  is $N T_0$, where $T_0=2\pi /\omega$ is the laser field period, and $N$ is the number of cycles.} One can see  that the ionization yield oscillates with the number of cycles.  The period of oscillations increases with larger $a_0$, indicating that it is a nondipole effect,  connected with the nondipole drift in the laser propagation direction $z$. The modulation depth of the oscillations increases up to $a_0=0.2$, with further damping at larger $a_0$. The latter indicates that there is another effect involved in the process which counteracts and competes with the nondipole drift.

As Fig.~\ref{Fig-1} points out the nondipole character of the oscillating yield effect, one might expect that it vanishes in the dipole case. In Fig.~\ref{Fig-2}(d) we compare the relativistic calculation with the nondipole  and the dipole ones. Surprisingly, the yield oscillation exists also in the dipole case, however, with a significant difference in the period, amplitude and shape of the oscillations compared with the nondipole case.

One can  note [Figs.~\ref{Fig-1}, \ref{Fig-2}] that the nondipole character of the field leads to the increase of the ionization yield, which naturally is explained by the effect of the nondipole  drift. The latter results in the shift of the wave packet average coordinate from the atomic core at the end of the pulse, which hinders the recombination of the wave packet and enhances the ionization, because
in the stabilization regime it emerges mostly during switching-on and -off of the laser field \cite{Gavrila_2002}.
Fully relativistic treatment yields a slightly suppressing of the ionization, because the relativistic mass correction slightly hinder the drift.

\begin{figure}
    \centering\includegraphics[width=0.495\textwidth]{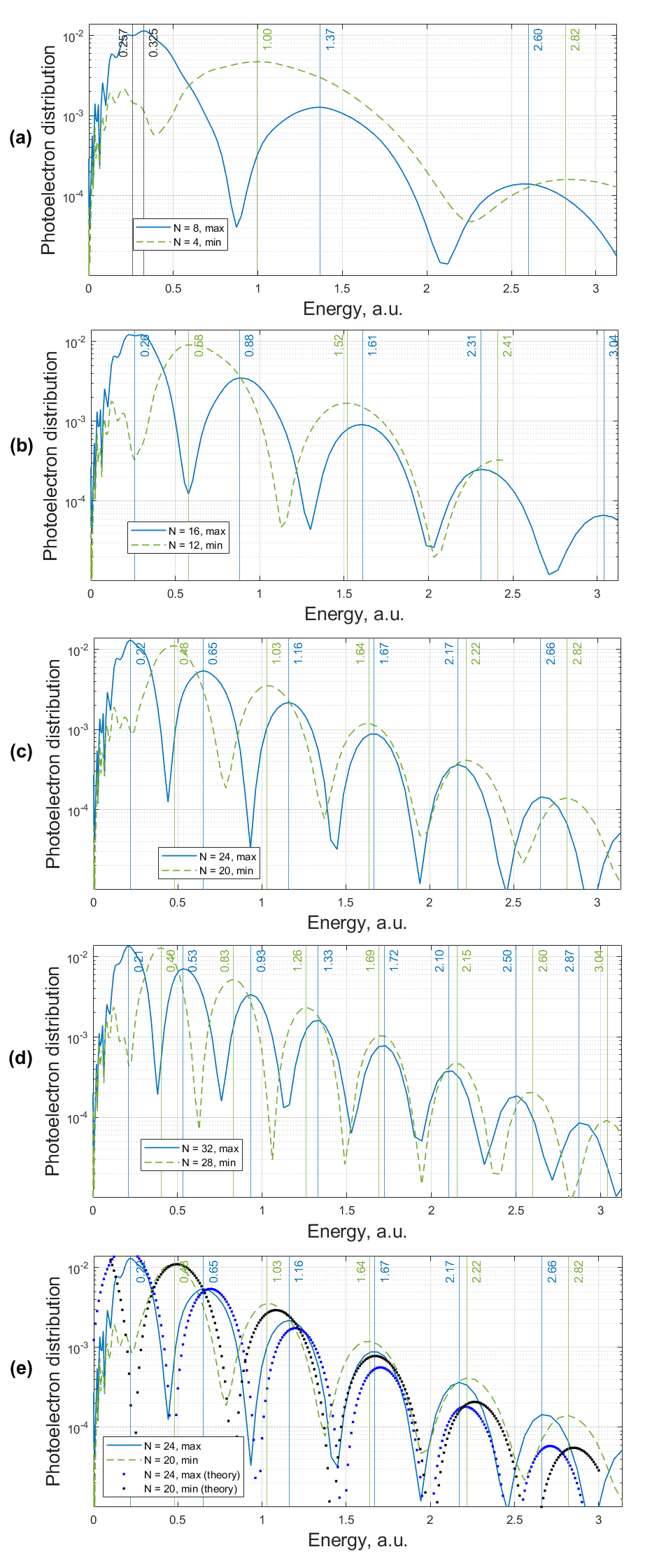}
    \caption{ \textbf{Photoelectron energy distributions (PED) in the dipole case}. PED near ZEP in the dipole calculations at different pulse durations, $a_0 = 0.13$, $\omega  = 14$~a.u.: (a)-(d) The number of cycles $N$ is indicated in the insets  and corresponds to the maxima and minima of the ionization probabilities of Fig.~\ref{Fig-2}; (e)  Comparison of the dynamic model Eq.~\eqref{SSS} (dotted line) with PEDs. }
    \label{Fig-3}
\end{figure}

\section{Dipole regime} \label{sec:dipole_regime}

Let us firstly examine the origin of the ionization yield  oscillation in the dipole  case  [Fig.~\ref{Fig-2}(d)]. In the dipole calculation, the ionization probability oscillates near an average value $w_i \approx 0.7 \%$. The oscillations seem to have a constant period ${\cal T} \approx 8 T_0\approx 3.59 ~\text{a.u.}$, independent of the total duration of the  pulse. For more information, we analyze  the photoelectron energy distribution (PED) in the dipole case in Fig.~\ref{Fig-3}. The chosen pulse durations  correspond to the local maxima or minima of the ionization probability   from Fig.~\ref{Fig-2}(d) (blue line). We show PED near the zero-energy peak (ZEP)  ($\varepsilon\ll \omega$), which contains   {most of the} ionization probability, see Table~\ref{table:1}.
\begin{table}[h!]
\begin{tabular}{|l|l|l|l|l|}
\hline
\diagbox{$a_0$}{$N$} & 1            & 5            & 12           & 20           \\ \hline
0.06   & 0.27         & 0.42         & 0.55         & 0.61         \\ \hline
0.09   & 0.07         & 0.10         & 0.13         & 0.16         \\ \hline
0.11   & 0.05 (0.04)  & 0.09 (0.05)  & 0.13 (0.06)  & 0.14 (0.08)  \\ \hline
0.13   & 0.03 (0.02) & 0.07  (0.03) & 0.09  (0.03) & 0.10  (0.05) \\ \hline
\end{tabular}
\caption{\textbf{The ratio of the  probability in the above-threshold ionization (ATI)  peaks with respect to the total ionization probability.}  The values are for the dipole case, and in brackets  for the nondipole case (where applicable). {  The nondipole TDSE and the
relativistic FW calculations give similar results for the applied parameters.} For the small $a_0$ value, the ATI channel dominates ionization, which  explains a steady growth in the Fig.~\ref{Fig-2}(a).}
\label{table:1}
\end{table}

\begin{figure}[b]
    \centering\includegraphics[width=0.5\textwidth]{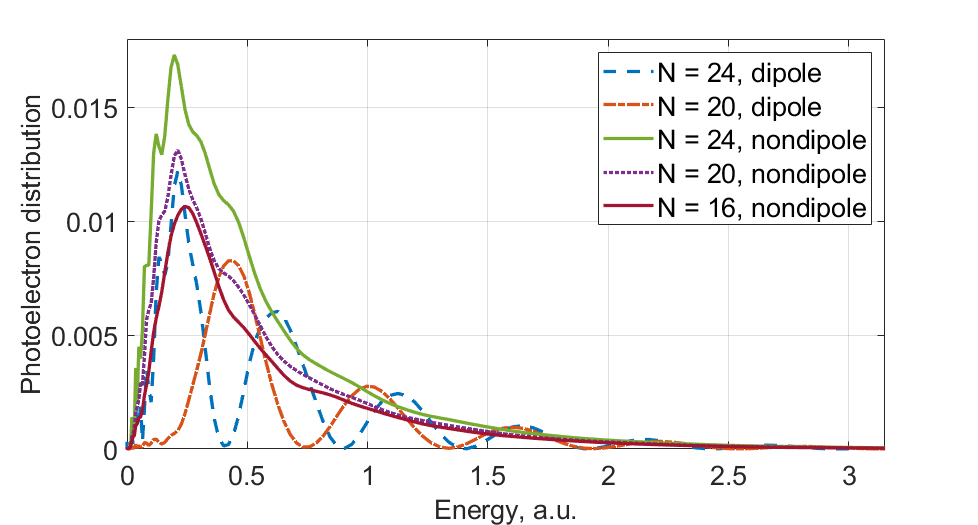}
    \caption{\textbf{Photoelectron energy distributions (PED) in the nondipole regime}. PED  for different pulse durations indicated in the inset: $a_0 = 0.11$. The dipole calculations are shown by the dashed lines, the nondipole  by the solid ones.}
    \label{Fig-4}
\end{figure}

\begin{figure*}
        \centering\includegraphics[width=0.70\textwidth]{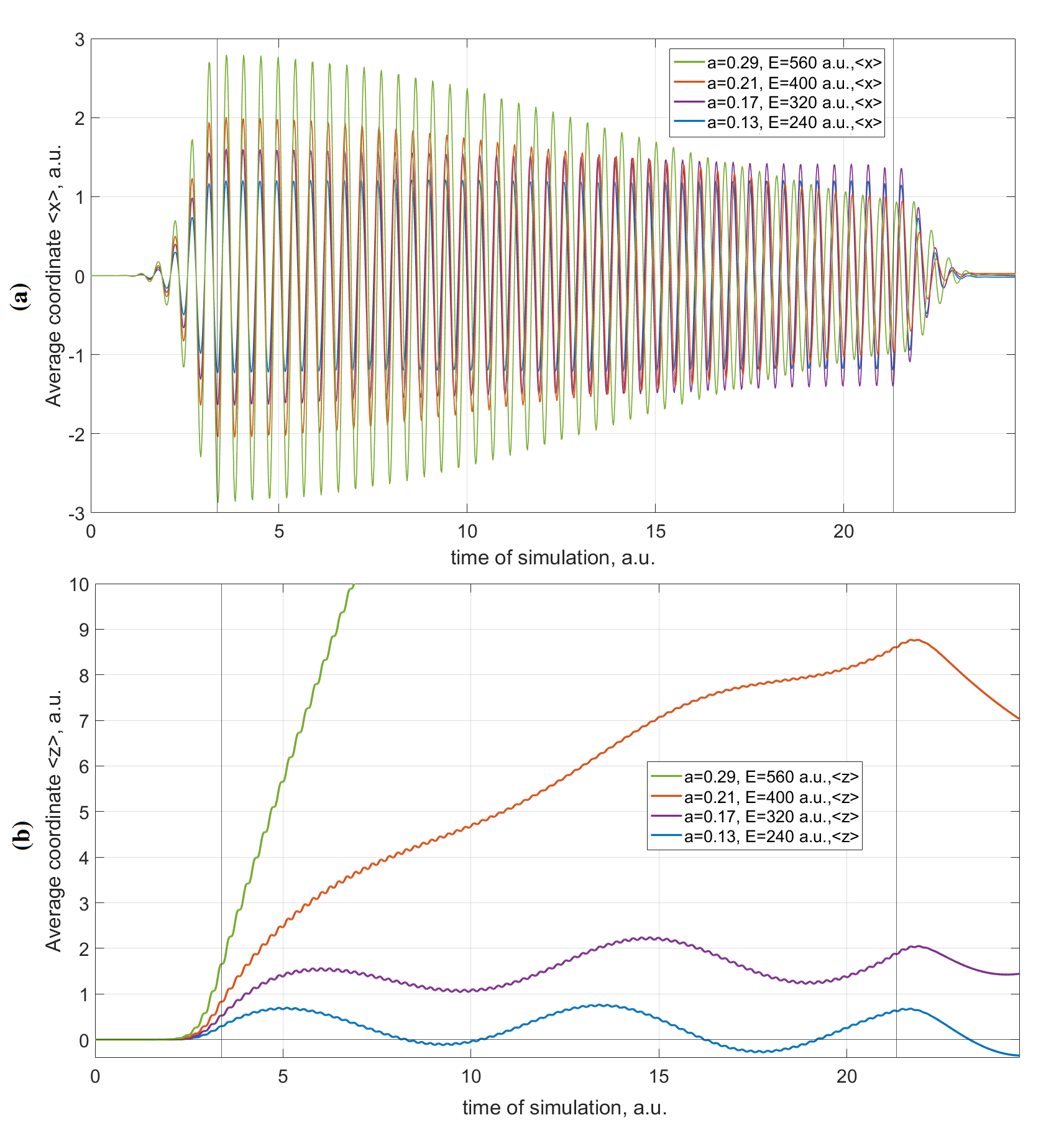}
    \caption{{ \textbf{  The expectation value for the electron coordinates during the interaction:} (a) The coordinate in the laser polarization direction  $\langle x(t) \rangle$, (b) the coordinate along the propagation direction $\langle z(t) \rangle$. Different lines represent different $a_0$ values indicated in the inset.  The frequency is $\omega = 14$ a.u. }}
    \label{Fig-5}
\end{figure*}

Firstly, we  note in Fig.~\ref{Fig-3} that the spectral density vanishes at the continuum threshold, $\varepsilon = 0$, which is in accordance with the PED scaling at the threshold at high laser frequencies $\omega>I_p $~\cite{Azizi_2025}. {  In our numerical calculation this property is accommodated by the fact that $\hat{\psi}( \mathbf{p} = 0) =   \int_{-\infty}^{+\infty} \psi(\mathbf{r} ) d^2\mathbf{r}\approx 0$ with a high accuracy.}

{  To explain the ionization yield oscillation  mentioned above, we take into account that the main ionization yield comes from the ZEP of the spectrum. Using  the dynamic interference picture, known from previous studies \cite{Toyota_2007,Toyota_2008, Demekhin_2012,Jiang_2018,Wang_2018, Geng_2021}, we put forward a parametrization scheme to describe the  numerically calculated PED. From the dynamic interference theory we know that the interference of the contributions to the ionization wave packet initiated at the trailing and falling edges of the laser pulse induces modulation of the PED, in particular, the modulation of ZEP. Rather than to develop a new microscopic theory of dynamic interference, we fit the numerically calculated spectra [Fig.~\ref{Fig-3}] to the parametrization of the ionization wave packet  chosen according to the main idea of the dynamic interference as follows
\begin{equation}
\label{yield}
    f(t) = f_0(t) + f_0(t - T)e^{-i\phi },
\end{equation}
introducing the parameters $T$ and $\phi$, as well as the  function $f_1(t)$. The most nontrivial parameter is the phase $\phi$. These parameters and the function $f_1(t)$  have been deduced from the numerically calculated PED spectra.
The concrete form of the ionization amplitude $f(t)$ is obtained by inspecting the envelope of the numerically calculated PED: $S(\varepsilon) \sim e^{-(const)\,\varepsilon}$.
By measuring the incline in the numerical PED, and using Fourier transform $f_0(t) \sim \mathcal{F}^{-1} \left\{e^{-\varepsilon\tau } \right\}= 1/[1+(t/\tau)^2]$,  we determine  $const$,  which as expected corresponds to the turn on time  of the laser pulse $\tau$:
\begin{eqnarray}
\label{SSS}
        S(\varepsilon) = e^{-\varepsilon\tau} \left[1+e^{i(\varepsilon T-\phi)} \right].
       \end{eqnarray}
The latter, we fit to the numerically observed PED to obtain the correct positions of the modulation peaks in ZEP.
From the analysis of the spectrum, we identify that $T$ corresponds to  $T = NT_0 + 2\tau$, and $T_0 = 2\pi/\omega$ is  the laser period.
Further, using the position of the peaks in the numerically calculated spectra
\begin{eqnarray}
\varepsilon_n=\Omega+(2\pi /T)n,
\end{eqnarray}
$n \in \mathbb{Z}$, see Fig.~\ref{Fig-3}(e), we are  able to recover the nontrivial phase $\phi = \Omega T$, with $\Omega \approx 8 T_0 \approx 3.59$~a.u., according to  Fig.~\ref{Fig-2}. The phase $\phi$ describes the following feature observed in PED: the position of the first nonzero modulation peak within the envelope of ZEP structure periodically oscillates with the laser pulse duration (the period of oscillation is ${\cal T}=2\pi/\Omega$). The oscillation of the PED modulation pattern results in the periodic oscillation of the integrated yield of the ZEP, as the most contributing peaks move to the damped region of the envelope, and the ionization at the threshold  $\varepsilon=0$ is suppressed. } Note that  $n=0$ corresponds to a high-order peak in PED (it is fixed at any pulse duration), while the first peak in the PED in the case of minimum yield is obtained at $n=-T/{\cal T}$, with $\varepsilon_n\approx 0$ (this threshold peak is suppressed).

The origin of the periodic term  $e^{-i\Omega T }$ in the ionization yield of Eq.~(\ref{SSS}) apparently  stems  from the bound state periodic dynamics in the KH potential during the interaction, described in Ref.~\cite{Aynul_2025}, with the parameter ${\cal T}$ {related to the energy difference of the populated KH states.}

 \section{Nondipole regime} \label{sec:nondip}
 \subsection{General features}

The transition from the dipole to the nondipole regime at increasing $a_0$  is illustrated in Fig.~\ref{Fig-2}(a-d) for the ionization yield. In the nondipole regime, there are also the ionization yield oscillations with respect to the interaction time, which have larger amplitude and longer periodicity. Also, the nondipole effect doubles the ionization probability,  $w_i \approx 1.4 \%$,  for the case $a_0 = 0.13$.  For the lowest presented $a_0 = 0.06$, we see that the ionization probabilities for the dipole and nondipole case follow the same pattern with the pulse duration. For the intermediate $a_0 = 0.09$, the plots are different by a constant for the short laser pulses $N < 10$, but the deviation becomes nontrivial for $N > 18$. At $a_0 = 0.11$, the dipole and nondipole behaviors follow different pattern.

Increasing the parameter $a_0$ further up to $0.29$ [Fig.~\ref{Fig-2}], we observe  a dramatic change of the ionization probability comparing to the dipole case. While at $a_0 = 0.13$, the yield behavior is still similar to its dipole counterpart, at $a_0 = 0.17$ it starts to exhibit both oscillations and a smooth increase, which is more pronounced at $a_0 = 0.21$.
At $a_0 = 0.29$,  we see a change of the pattern again. There are no more large oscillations and the ionization gradually increases with respect to the pulse duration up to the saturation.

The PED in the nondipole case, see Fig.~\ref{Fig-4}, in contrast to the dipole case, is dominated by a single low energy broad distribution, coinciding with the bandwidth of the dipole PED $\Delta\varepsilon\sim 2\pi/\tau$, without significant modulation due to the dynamic interference. Thus,  the dynamic interference cannot be the main cause for the yield oscillation in the nondipole regime.

To analyze the nondipole features of the dynamics, we inspect the expectation values of the coordinates for the electron wave packet in the continuum during the interaction. The typical trajectories via the expectation value of the coordinates in the laser polarization direction $\langle x(t) \rangle$, and the propagation direction $\langle z(t) \rangle$ for  $a_0 = 0.13 - 0.29$ are shown in Fig.~\ref{Fig-5}.
The main difference of $\langle x(t) \rangle$ from the dipole regime becomes visible after $a_0 \gtrsim 0.17$, namely the amplitude of the oscillations starts gradually decreasing. This is due to the extended size of the electron quantum wave packet, when its different parts  are exposed to the different values of the nondipole field,  tending to decrease oscillations of an electron in space with time.

For the expectation value $ \langle z(t) \rangle$, the situation is less trivial. In case of the absence of the atomic potential, the electron would experience a magnetically induced drift in the laser propagation $z$ direction  with an average speed $v_z(t) = A_0^2(t)/(4c)$ \cite{Salamin_1996}.  However, the atomic potential counteracts  the laser induced drift, which results in the generation of the electron slow oscillations along the $z$ coordinate, with a frequency much smaller than the laser frequency. At larger $a_0$, the drift compensation fails and  the drift component in $ \langle z(t) \rangle$ starts to dominate, see the case  $a_0=0.29$ in Fig.~\ref{Fig-5}.

\begin{figure}
\centering
\includegraphics[width=0.5\textwidth]{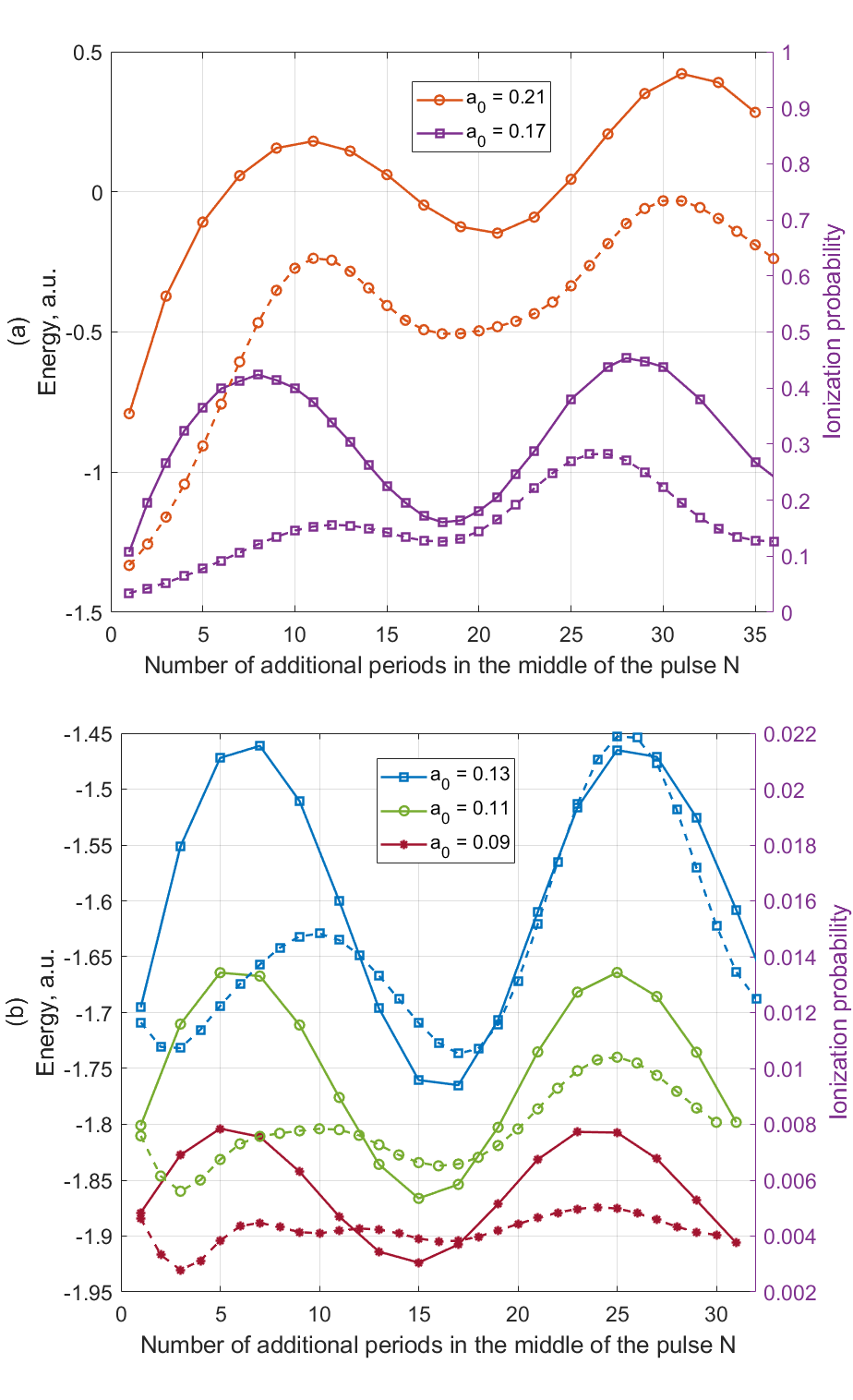}
\caption{{  \textbf{The electron energy after the interaction as a characteristic parameter of the ionization yield.} The expectation value of the electron energy $\langle \varepsilon \rangle$ of the electron wave packet $t=10 ~\text{a.u.} $ after the  interaction vs the laser pulse duration (solid lines):  (a) $a_0=0.17$ (violet),  $a_0=0.21$ (orange) , (b) $a_0=0.09$ (red), $a_0=0.11$ (green), $a_0=0.13$ (blue).  The corresponding ionization yields are shown in the panels as dashed lines.  } }
\label{Fig-6}
\end{figure}

The periodic oscillation of the expectation value of $\langle z(t) \rangle$ for the electron  wave packet during the interaction results in the oscillation of $\langle z \rangle$, as well as the average energy {  $\langle \varepsilon\rangle=\langle \mathbf{p}^2\rangle/2+\langle V(z) \rangle$} at the switching-off of the laser field. In the stabilization regime when the main ionization takes place at the switching-on and -off of the laser pulse, the final coordinate and,  hence,  the energy oscillation yield  a different probability of capturing the electron by the atomic field, and  result in the oscillation of the ionization yield, see Fig.~\ref{Fig-6}. The latter shows that the maximum $\langle \varepsilon  \rangle$  is correlated with the maximum ionization yield, and it can serve as a characteristic parameter for the ionization yield.

 \begin{figure} [b]
    \centering\includegraphics[width=0.45\textwidth]{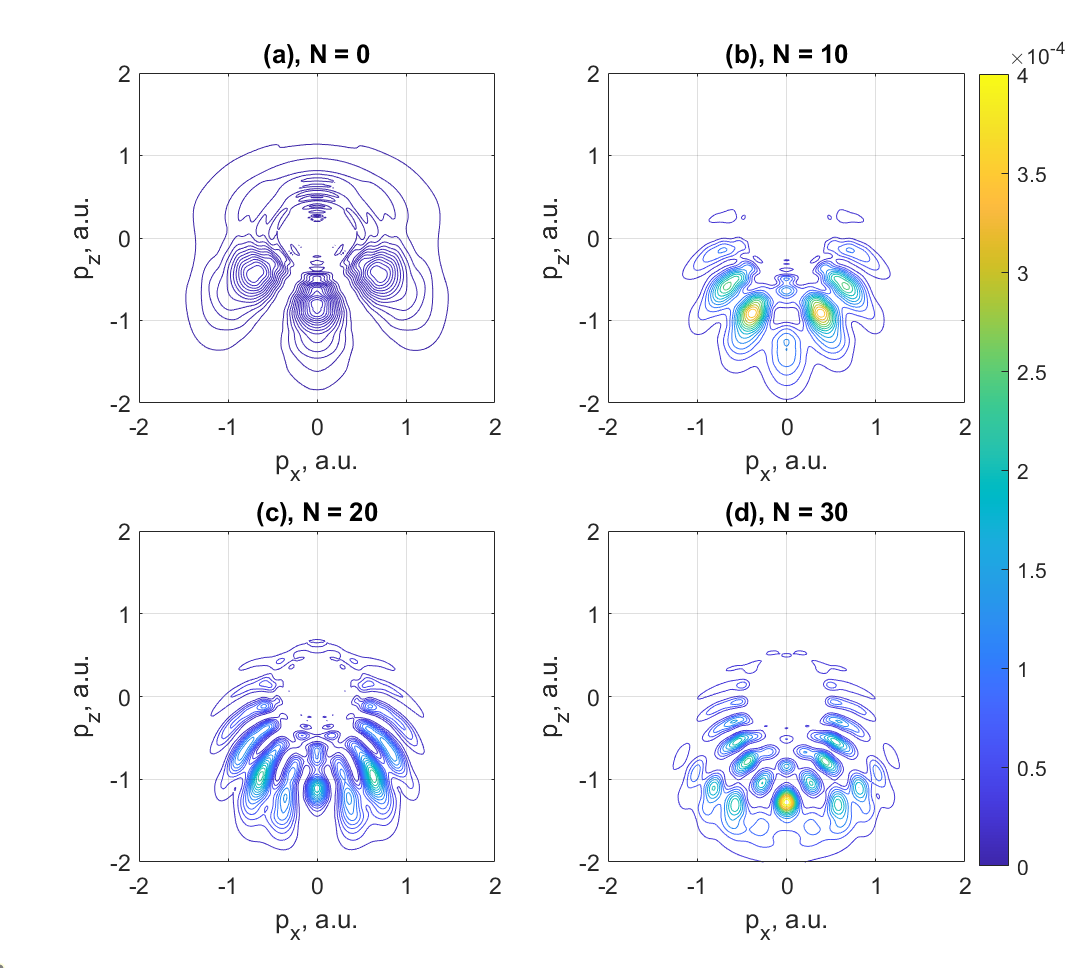}
    \caption{\textbf{Photoelectron momentum distributions (PMD)
        for different durations of the pulse.} (a) $N = 0$, (b) $N = 10$, (c) $N = 20$, (d) $N = 30$. The parameters are: $a_0 = 0.21, ~\omega = 14 ~\text{a.u.}$, corresponding to the red line in Fig.\ref{Fig-1} and Fig.\ref{Fig-5}.}
    \label{Fig-7}
\end{figure}

\begin{figure}[b]
         \centering
         \includegraphics[width=0.4\textwidth]{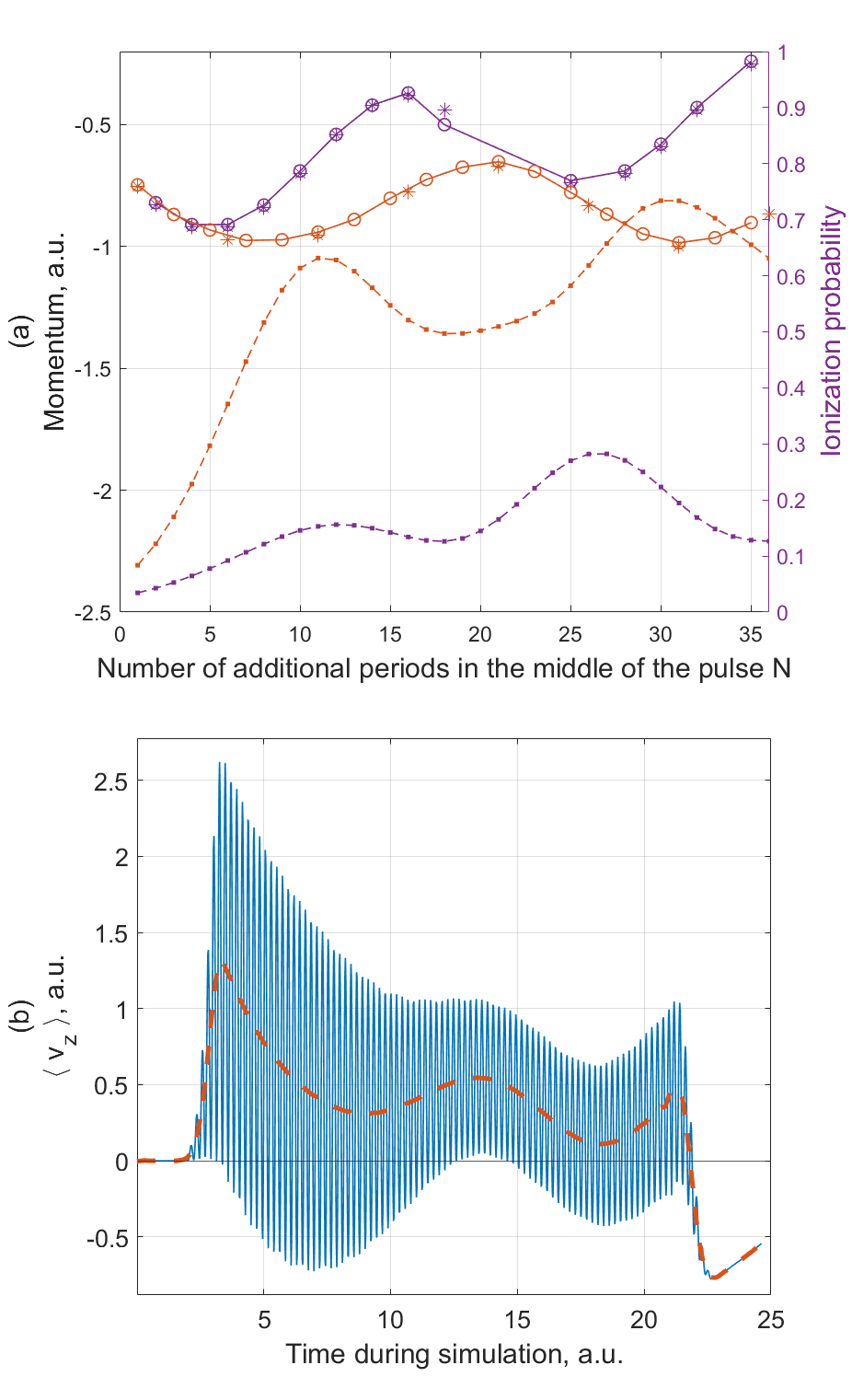}
         \caption{\textbf{ Correlation between the Coulomb momentum transfer (CMT) during the interaction and the ionization yield in the nondipole regime.}   (a)  CMT (stars), $\langle p_z  \rangle$ (solid lines) and the ionization yield (dashed lines) vs the pulse duration, (violet) $a_0=0.17$, (red) $a_0=0.21$, the corresponding ionization yields are shown as dashed lines; (b)  Average velocity $\langle v_z(t) \rangle$ of the electron wave packet vs the interaction time: Blue line is  $\langle v_z(t) \rangle$, the red line is an averaged  $\langle v_z(t) \rangle$ over  the fast oscillations, $a_0 = 0.21,  ~\omega = 14 ~\text{a.u.}$. }
    \label{Fig-8}
\end{figure}

\subsection{Nondipole regime: Coulomb momentum transfer}\label{sec:CMT}

It is known that the Coulomb effects play an essential role for the strong field ionization in the stabilization regime \cite{Foerre_2006}. The photoelectron momentum distribution (PMD)  in this case shows an additional lobe into the direction opposite to the laser propagation direction, see Fig.~\ref{Fig-7} (the appearance of the interference structures in PMD is discussed in Ref.~\cite{Geng_2021}). This counterintuitive behavior is the result of the interplay between the electromagnetic and Coulomb forces. Due to the nondipole drift the electron spends more time during the interaction at $z>0$ and experience the Coulomb momentum transfer (CMT)  in the opposite direction. Moreover, the classical  trajectories of ionized electrons which contribute to the anomalous nondipole lobe in PMD will be captured in the case of  the dipole interaction. The larger the Coulomb momentum transfer during the interaction is, the larger will be the contribution into the PMD anomalous lobe. Therefore, we anticipate that the  CMT during the interaction will provide an additional characteristic parameter for the ionization yield in the nondipole regime $a_0\gtrsim 0.1$. We calculate CMT during the interaction
\begin{equation}
    \label{eq:newton}
     \langle\mathbf{p}^C(t)\rangle =-\int_{-\infty}^t dt \langle \boldsymbol{\nabla} V(r) \rangle,
     \end{equation}
with the expectation value of the Coulomb force $\langle -\boldsymbol{\nabla} V(r) \rangle$. The total values of CMT $\langle p^C_z  \rangle$ after the end of the interaction for  $a_0=0.17$ and $a_0=0.21$ are presented in Fig.~\ref{Fig-8}(a). For the both cases, the expectation value of the momenta of the final electron wave packet $\langle p_z  \rangle$ and the Coulomb induced momenta Eq.~\eqref{eq:newton} nearly coincide. At $a_0=0.21$ there is a good correlation to the ionization yield. At $a_0=0.17$ the correlation is mostly qualitative. This is because the CMT of Eq.~(\ref{eq:newton}) is calculated via the total wave function of the electron, but the part of it recombines to the ground state after the interaction. Thus, we can take the total CMT $\langle p^C_z \rangle$ as a characteristic parameter for the ionization in the nondipole regime when the ionization yield is significant and is mostly determined by the ZEP.

In the case of $a_0=0.21$, from one side the CMT shows well-defined oscillations with respect to the interaction time [Fig.~\ref{Fig-8}(a)], but from another side $\langle z(t) \rangle$ increases almost monotonously with the interaction time [Fig.~\ref{Fig-5}], which would raise a question on the source of CMT. It is right that $\langle z(t) \rangle$ (and the distance from the core) increases with the interaction time due to the nondipole drift, however, the dynamics includes additionally slow oscillation induced by the Coulomb force. Due to the latter the electron wave packet is slowed down at the turning points of the Coulomb induced orbiting, {remains a} long time at the same position, and acquires a large CMT. This is confirmed by Fig.~\ref{Fig-8}(b) showing that the CMT is correlated with the slowing down of $\langle v_z \rangle$.

\subsection{ Coulomb oscillations of the electron wave packet in continuum} \label{sec:orbiting}

We have seen, that the ionization yield oscillation in the nondipole regime depends on the laser pulse duration and is related to the periodic oscillatory dynamics of the electron wave packet in the laser propagation direction [Fig.~\ref{Fig-5}]. The latter results in the oscillating energy and CMT at the end of the laser pulse, which finally has an impact on the ionization yield. Here, we examine more closely the slow quasiperiodic Coulomb oscillation dynamics, averaged over the fast laser driven oscillations along $x$, which arises as a result of the competition between the laser induced drift along $z$ coordinate and the Coulomb force.
 To this end, we invoke Ehrenfest equation for the expectation values:
\begin{eqnarray}
\frac{d \langle\mathbf{v}(t)\rangle}{dt}=-\langle E( \mathbf{r},t)\rangle-\frac{1}{c}\langle\mathbf{v}(t)\times \mathbf{B}( \mathbf{r},t)\rangle-\langle \boldsymbol{\nabla} V(r) \rangle,
\label{eq:Ehrenfest}
\end{eqnarray}
 { which reads in the components as
\begin{eqnarray}
\label{eq:Ehrenfest2}
\frac{d^2 \langle x(t)\rangle}{dt^2}&=&-\left(1- \frac{\langle v_z\rangle}{c}\right)E_0\sin\omega t-\langle \partial_x V (x,z)\rangle\\
\frac{d^2  \langle z(t)\rangle}{dt^2}&=&-\frac{\langle v_x\rangle}{c}E_0\sin\omega t-\langle\partial_z V (x,z)\rangle,\nonumber
\end{eqnarray}
where $V(x,z)=Z/\sqrt{x^2+z^2+a_s^2}$, with the nuclear charge $Z=2$, and the  soft core parameter $a_s=\sqrt{3}/Z$ a.u. We have taken into account the nondipolness of the interaction including the laser magnetic field.

In order to solve Eq.~\eqref{eq:Ehrenfest2} as a classical equation, we employed  approximations for the expectation value of the Coulomb force $-\langle \partial_x V(x,z) \rangle$. We apply different approximations at $a_0 \ll 1$ (cf. the cases $a_0\lesssim 0.13$ in Fig.~\ref{Fig-5}), and at $ a_0 \sim 0.2$ (cf. $a_0=0.17$ and $a_0=0.21$ in Fig.~\ref{Fig-5}).

In the first case the typical size of the coordinate oscillation is less than the initial electron wave packet size as well as that  throughout the interaction. Then, we introduce an effective atomic potential by means of smoothing the original soft-Coulomb atomic potential with an original ground state $\psi_0(\mathbf{r})$ as
\begin{eqnarray}
    \langle V (\mathbf{r})\rangle = \int_0^{\infty} V(\mathbf{r}') |\psi_0(\mathbf{r}-\mathbf{r}')|^2 d\mathbf{r}'.
    \label{QC}
\end{eqnarray}
We estimate the condition for this regime as $\alpha_0\sim a_s$, which yields
\begin{eqnarray}
a_0\sim \frac{\sqrt{3}\omega}{cZ}\approx 0.1,
\end{eqnarray}
for our parameters $Z=2$ and $\omega=14$.

In the opposite case, the oscillation typical size along $z$ axis is larger than the initial electron wave packet size [Fig.~\ref{Fig-5}]. Then, the electron wave packet is relatively compact during the interaction and we use the approximation
\begin{eqnarray}
-\langle \partial_x V(x,z) \rangle \approx - \partial_x V(\langle x \rangle , \langle z \rangle )
       \label{ClC}
\end{eqnarray}
We calculate the numerical solution of Eq.~(\ref{eq:Ehrenfest2}) with either the effective potential of Eq.~(\ref{QC}), or with Eq.~(\ref{ClC}) and derive  the corresponding oscillation periods, see blue and magenta lines in Fig.~\ref{Fig-9}.
\begin{figure}
    \centering\includegraphics[width=0.48\textwidth]{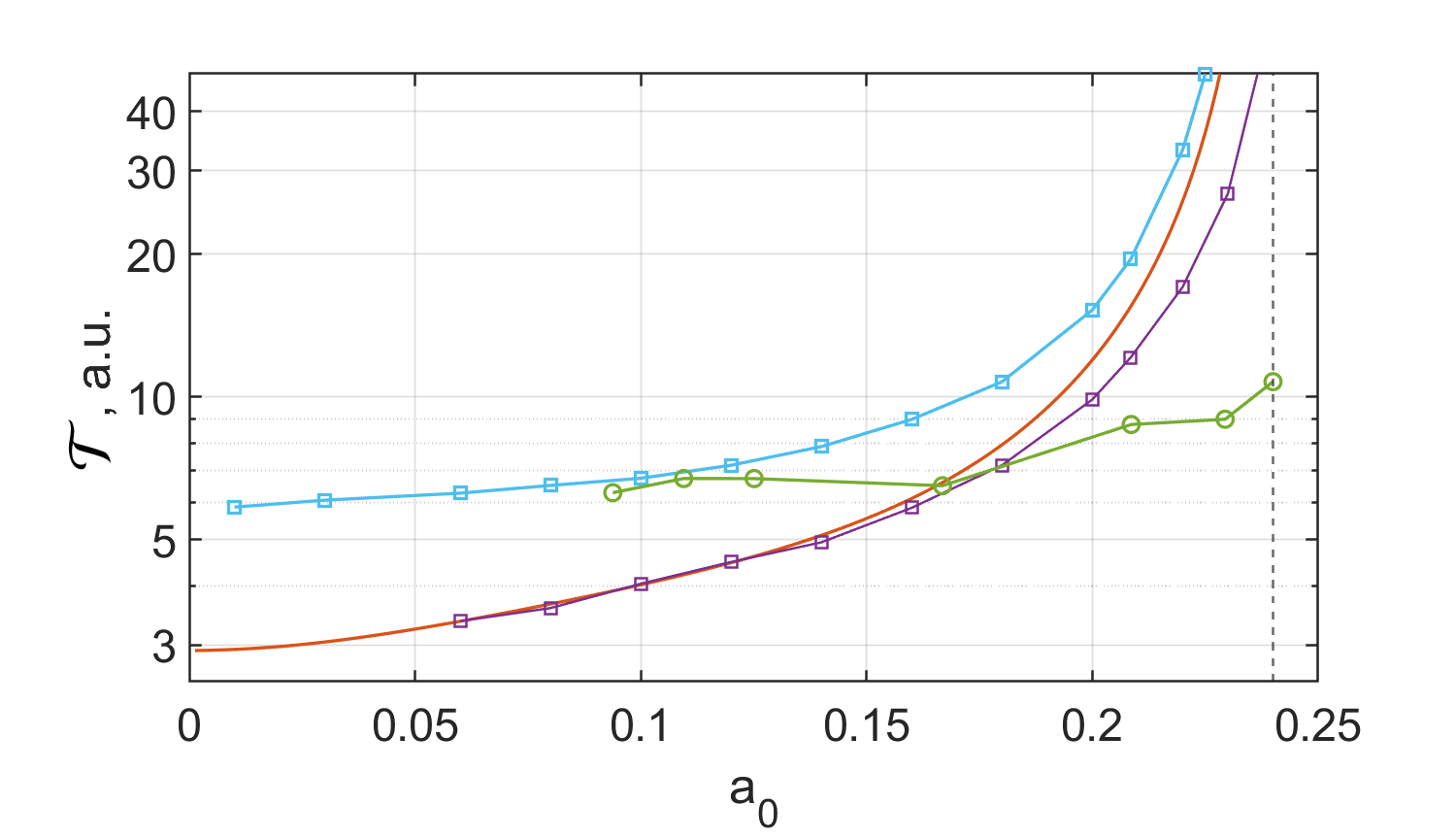}
    \caption{ \textbf{Period of the slow Coulomb induced oscillations during the interaction.} The period  vs the field parameter $a_0$: (blue solid line, squares)  the period  extracted from the exact solution of Eq.~\eqref{eq:Ehrenfest2} using the quantum average of the Coulomb potential [Eq.~(\ref{QC})], (magenta solid line, squares) the period in the case of the atomic potential of Eq.~(\ref{ClC}); (red solid line) the period  extracted from the solution of the  approximate Eq.~\eqref{eq:V_osc}. Green line with circles -- a period of oscillations of the quantum ionization probability. Dashed black line -- the threshold $a_0^{\rm (th)}$.
       }
    \label{Fig-9}
\end{figure}

To obtain the scaling for the slow oscillation parameters, we apply an analytical analysis, seeking for an analytical approximate solution for the coordinate expectation values, representing it as slow and fast motions:
\begin{eqnarray}
\langle x(t)\rangle =  {\cal X}(t)+\  \xi(t),\,\,\,\,\langle z(t)\rangle = {\cal Z}  (t)+\zeta(t),\label{XZ}
\end{eqnarray}
where fast motion $\xi(t), \zeta(t)$ is characterized by the time scale of the laser period $T_0$.
We insert the ansatz Eq.~(\ref{XZ}) into Eq.~(\ref{eq:Ehrenfest2}), and average over the fast oscillations.
We assume that the laser field dominates the Coulomb one $E_0\gg E_a$, with the atomic field $E_a=Z/a_s^2$ then the fast oscillations are described only by the laser field
\begin{eqnarray}
\label{eq:Ehrenfest3}
\xi (t)&=&\alpha_0\sin\omega t ,\\
\zeta(t)&=&\frac{ca_0^2}{4\omega } \sin 2\omega t.
\end{eqnarray}
The laser induced nondipole drift ${\cal Z}_L  =v_{zd}t$, with $v_{zd}=c a_0^2/(4 + a_0^2)$ \cite{Salamin_1996},  is included in the {slow dynamics. Since $\ddot{\cal Z}_L = 0$, the electron slow motion as a result of competition of the laser induced drift and the Coulomb attraction is described as
\begin{eqnarray}
\label{eq:V_osc}
 \ddot{{\cal Z}} =  -\partial_{\cal Z}\overline{ V  }({\cal Z}, \alpha_0),
 \end{eqnarray}
where the laser induced drift is accounted for via the initial condition $\dot{{\cal Z}}(0) = v_{zd}$. Here, the overline indicates averaging over the fast oscillations.  $\overline{ V  }({\cal Z}, \alpha_0)$  depends on the distance from the potential center $r$, which can be approximated as $r^2=({\cal X}+\xi)^2+({\cal Z}+\zeta)^2+a_s^2\approx \xi^2+{\cal Z}^2+a_s^2$, taking into account that  $|{\cal X}|\ll |\xi(t) |$, and $|\zeta(t)|\ll |{\cal Z}|$, which is in accordance with the results of Fig.~\ref{Fig-5}.
The averaged Coulomb potential can be presented as
\begin{eqnarray}
\label{eq:KHpotentail}
\overline{V  }({\cal Z}, \alpha_0)&\approx &\frac{1}{T_0}\int_{0}^{T_0} dt'\frac{ Z}{\sqrt{\xi(t')^2+{\cal Z}^2+a_s^2} },\nonumber \\
 & = & -  \dfrac{4Z}{\sqrt{a_s^2+{\cal Z}^2}} K \left(-\dfrac{\alpha_0^2}{a_s^2 + {\cal Z}^2} \right)
\end{eqnarray}
where $K(.)$ is the complete  elliptic integral of the first kind.

In the limit $a_0 \ll 1$, the laser drift is not large $|{\cal Z}_L|\ll|{\cal Z}|\ll \alpha_0$  and the KH force can be expanded with respect to the small parameter $|{\cal Z}_C|/\alpha_0$: $-\partial_{\cal Z} \overline{V}({\cal Z}) \approx - [Z/(a_s^2+\alpha_0^2)^{3/2}]{\cal Z}$, which results in the harmonic slow oscillations
\begin{equation}
    \ddot{{\cal Z}}  +\Omega^2 {\cal Z}=0.
\end{equation}
{ Here,
\begin{eqnarray}\label{period}
\Omega=\frac{\sqrt{Z}}{(a_s^2+\alpha_0^2)^{3/4}},
\end{eqnarray}
which provides the scaling of the oscillation frequency with respect to the laser and atom parameters.}

 The  period of the slow oscillation ${\cal T}=2\pi/\Omega$ via the solution of simple approximate   Eq.~(\ref{eq:V_osc}) (red line), and the exact numerical solution of Eq.~\eqref{eq:Ehrenfest2} with the approximation Eq.~(\ref{ClC}) (magenta line) are shown in Fig.~\ref{Fig-9}.  Both  results coincide quantitatively for the case $a_0 <0.13$, and exhibit the same trend for the general case $a_0 \rightarrow 0.25$, which confirms the accuracy of the  approximation in Eq.~\eqref{eq:V_osc}.

The oscillatory dynamical picture  in the KH potential is valid  when the drift distance during one slow oscillation $\sim v_{zd}/\Omega$ does not exceed the Coulomb induced oscillation amplitude $z_C $: $v_{zd}/\Omega \lesssim z_C$.
We estimate $z_C $ from the energy conservation at Coulomb oscillation in the KH potential: $ v_{zd}^2/2= Z/\sqrt{a_s^2+\alpha_0^2+z_C^2}$,
approximate in this regime $\Omega\approx Z/\alpha_0^{3/2}$, and $z_C\approx 2Z/v_{zd}^2$, and  obtain the condition  for the oscillatory dynamical picture:
 \begin{eqnarray}
 \label{a0_1}
  a_0  \lesssim   \left( \frac{16\cdot2^{2/3}Z\omega}{c^3}\right)^{1/5}\approx 0.2,
\end{eqnarray}
 for $\omega=14$, $Z=2$.

The periodicity of the ionization yield  disappears at larger $a_0$ (cf. the case of $a_0=0.29$ in Figs.~\ref{Fig-1} and \ref{Fig-5}) when the drift dominates the Coulomb force. We may estimate this threshold assuming the kinetic energy of the electron at switching-on the field $v_{zd}^2/2$ exceeds the Coulomb attraction estimated by $I_p=Z^2/2$, which results in the condition for the threshold when the   periodicity of the ionization yield  disappears
\begin{eqnarray}\label{a0_2}
a_0\gtrsim a_0^{\rm (th)}=\sqrt{\frac{4Z}{c}},
\end{eqnarray}
for our case $a_0^{\rm (th)} \approx 0.24$. {  Thus, the different regimes of the oscillatory dynamics of the ionization wave packet are delimited via Eqs.~(\ref{a0_1}) and (\ref{a0_2}).}
 \begin{figure}[b]
    \begin{center}

        \includegraphics[width=0.5\textwidth]{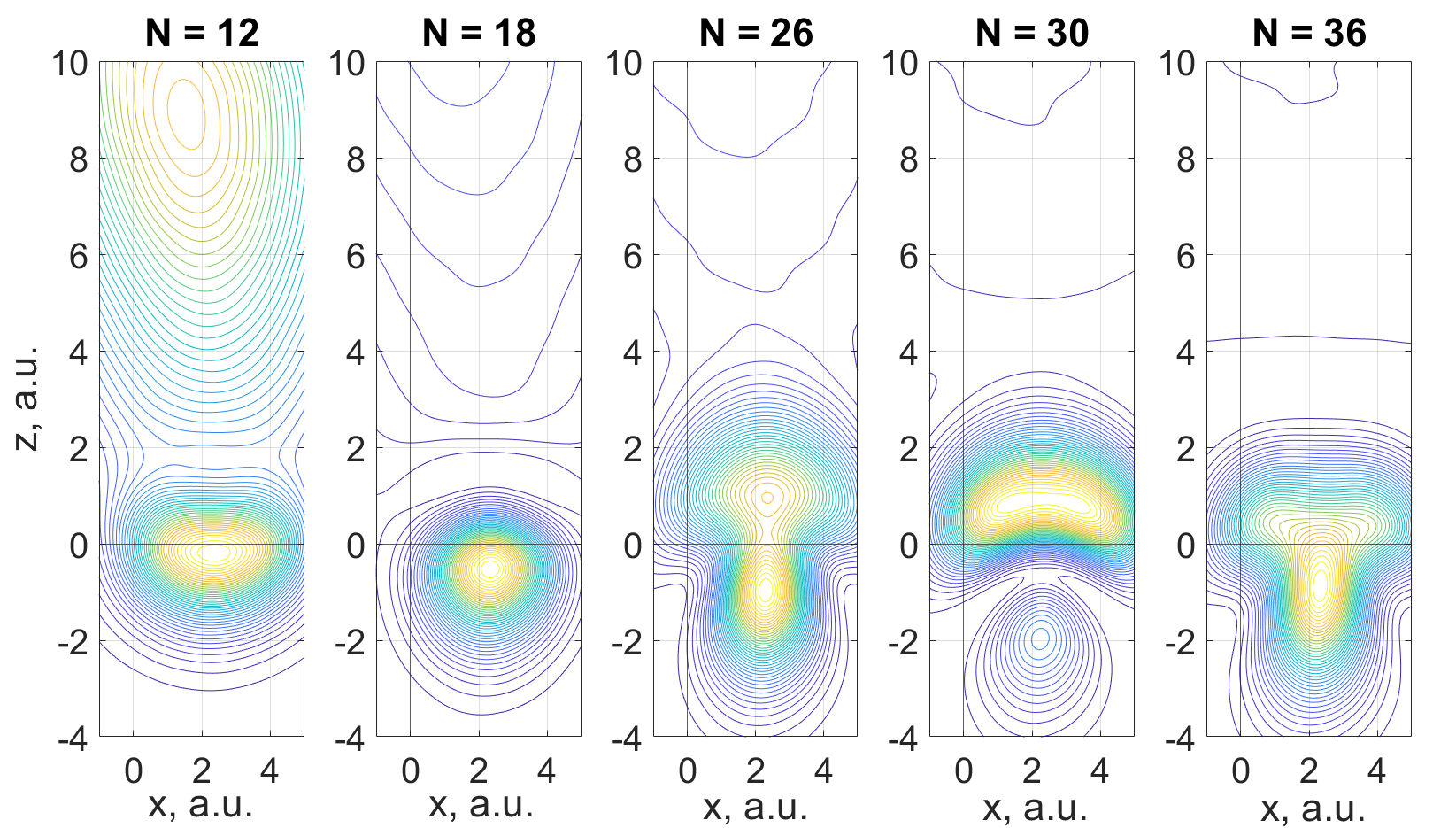}
         \caption{{  The electron wave packet dynamics during the interaction via the time-dependent probability density $|\psi|^2$, $a_0 = 0.24, E=460$ a.u.. The panel $N$ corresponds to the moment $NT_0 $   after the turn-on phase. The core of the slow oscillating part of the electron is located roughly between the upper and the bottom halves of the plane. At $N = 18$ the electron is in the bottom half of the plane. At $N = 26$ the electron transits from the bottom to the top and its center is located again at $z = 0$. At $N = 30$, the electron is in the upper half. At $N = 36$, the electron transits again via $z = 0$ line from the upper half to the bottom one. The time needed for such cycle is $(36-12)T_0 = 10.8$ a.u., which is in quantitative agreement with the measured quantum ionization period (see green line in Fig.~\ref{Fig-9}).}}
         \label{Fig-10}
\end{center}
\end{figure}

In turn, the oscillation period of the ionization yield is extracted directly from the quantum calculations [Fig.~\ref{Fig-1}] and is presented in Fig.~\ref{Fig-9} (green line). There is a qualitative accordance between the  oscillation periods of the ionization yield and that of the $\langle z\rangle$ expectation value   via  the classical estimate of the blue line for $a_0\lesssim 0.13$, and of the magenta line within the interval $0.15\lesssim a_0\lesssim  0.2$. At larger $a_0\gtrsim 0.2$,  the period of the ionization yield does not significantly grow as $a_0$ increases. It remains on the level, corresponding to the period of $\langle z\rangle$ oscillations via the classical estimate  for $a_0 \lesssim 0.2$. This stems from the fact, { that during the interaction the electron is not represented as a single wave packet,  see Fig.~\ref{Fig-10}.} In fact, one part of the electron wave packet always remains localized near the center of the Coulomb potential and experiences slow   oscillations described by the analysis of the Ehrenfest equation. This part of the electron wave packet is responsible for the ionization yield oscillation. The other part of the wave packet drifts in the positive $z$ direction, $z_d = v_{zd}T$, and is responsible for the gradual increase of the ionization yield with the duration of the pulse [Fig.\ref{Fig-5}]. After the end of the pulse, the shifted part of the electron is attracted back to the Coulomb center and diffracted by the atomic core, producing the spectrum in Fig.~\ref{Fig-7}. The larger $a_0$, the more pronounced is the contribution from the drifted part of the electron, which results in the decrease of the oscillation magnitude, making the yield oscillation hardly visible as $a_0 \rightarrow 0.29$ [Fig.\ref{Fig-5}].

For the $a_0$ values in the interval $(0.2,0.24)$ the oscillation period from the quantum calculations saturates, while the solution of the classical Eq.~(\ref{eq:V_osc}) shows an oscillation with an increasing period. The latter is possibly connected with the fact that the classical trajectories near the separatrix of the ionization threshold with a large oscillation period are actually ionized in the quantum treatment.

{  We emphasize the connection between the slow oscillatory motion of the wave packet and the modulation of the ionization yield.  The regime of interaction, depending on the value of $a_0$, is related to the relative probabilities of the  two parts of the wave packet shown in Fig.~\ref{Fig-10}.  When $a_0$ is not large fulfilling the condition of Eq.~(\ref{a0_1}), the regular oscillation of the trajectory and the ionization yield takes place with a period of Eq.~(\ref{period}). At a larger $a_0$ from the threshold value of Eq.~(\ref{a0_2}), the regular periodicity is destroyed.

\section{Pulse shape effects}\label{sec:pulse}
 \begin{figure}[b]
    \begin{center}
            \includegraphics[width=0.5\textwidth]{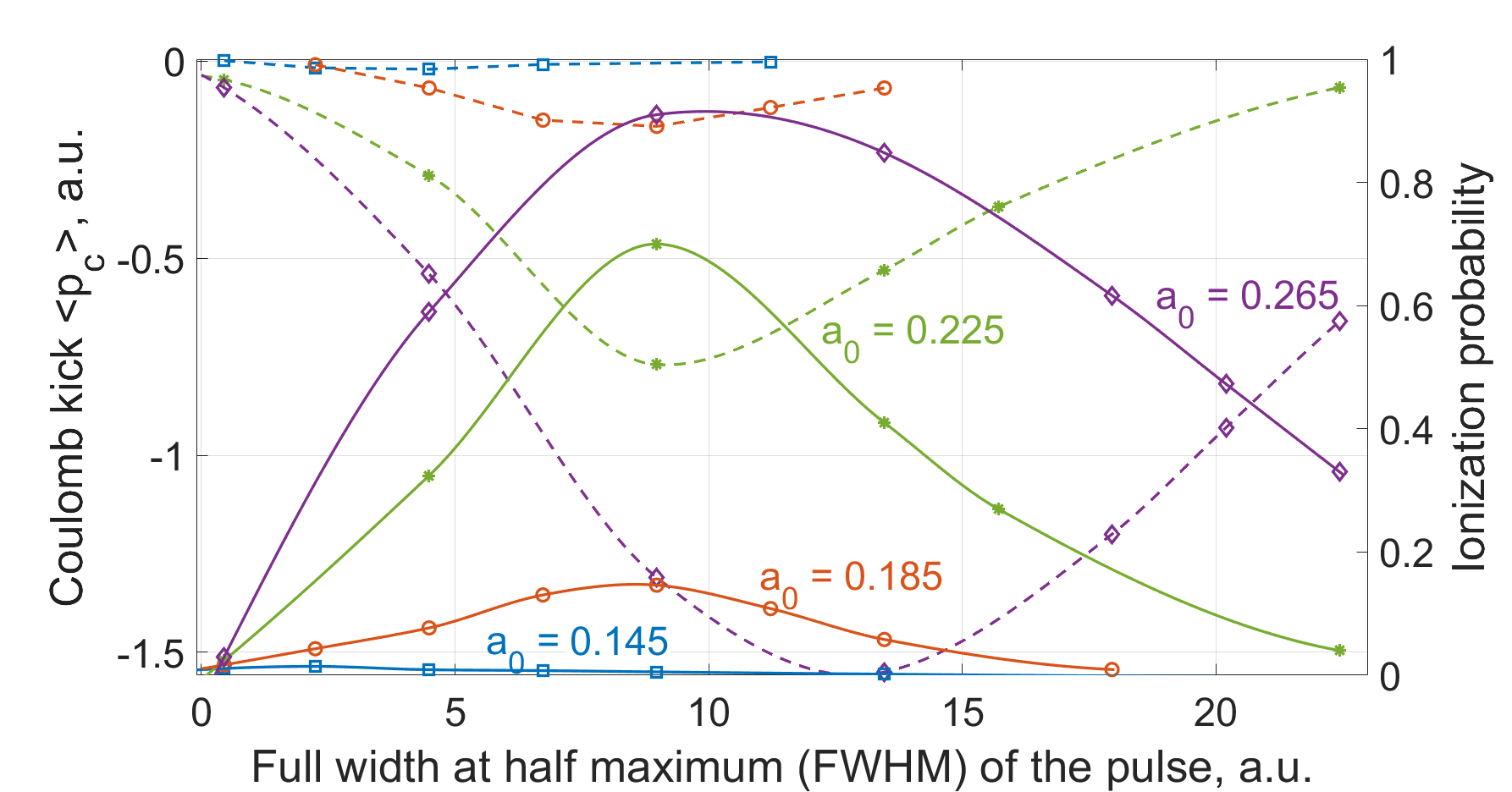}
    \caption{{  The ionization probability (solid line, right scale) and CMT (dashed line, left scale) vs the laser pulse duration in the nondipole regime: $Z=2$, $\omega = 14 ~\text{a.u.}$. The laser pulse is of a Gaussian form.} }\label{Fig-11}
   \end{center}
\end{figure}
\begin{figure}
\begin{center}
\includegraphics[width=0.4\textwidth]{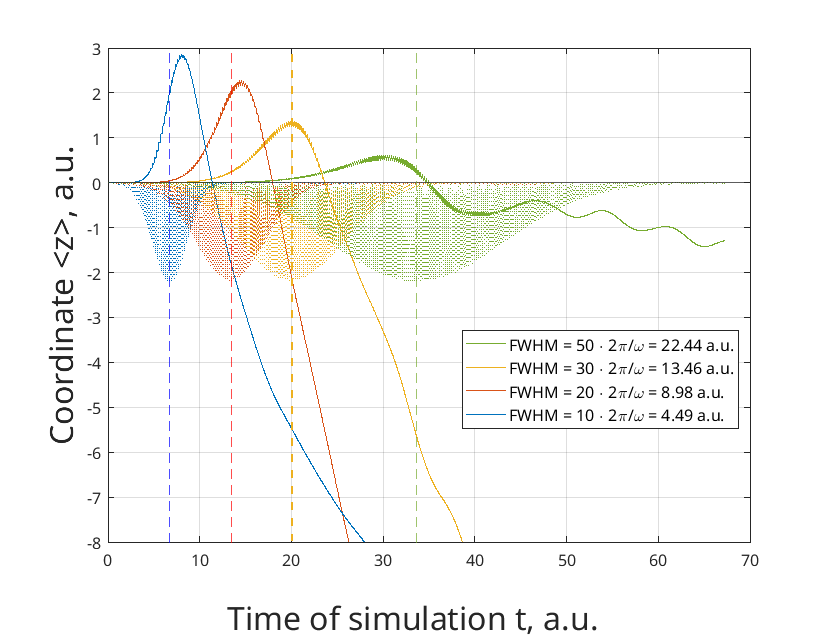}
\includegraphics[width=0.4\textwidth]{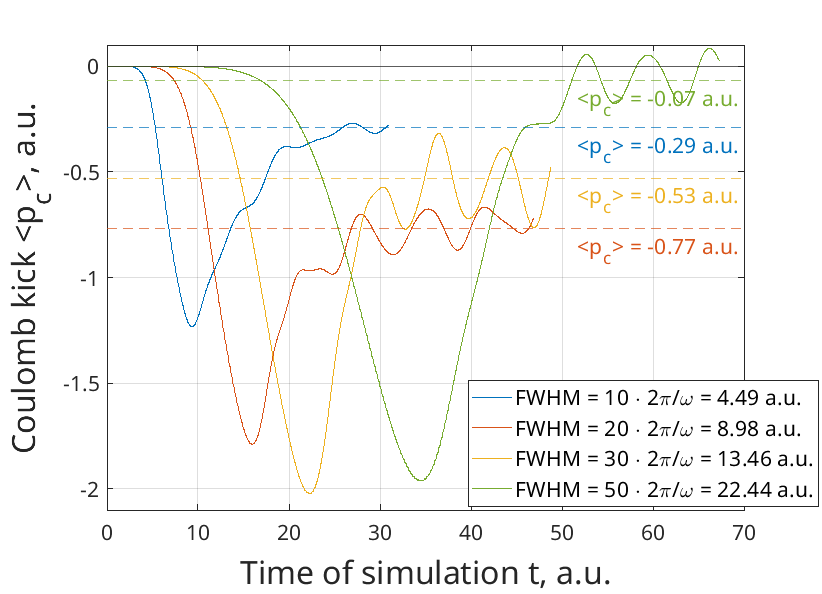}
    \caption{ { (top)  Trajectory of an electron $\langle z\rangle$ vs the time of simulation in a Gaussian pulse, $Z = 2, ~\omega = 14 ~\text{a.u.}, a_0 = 0.225,~ E=430$ a.u. The FWHMs are indicated in the inset.  Color coding is the same as in Fig.~\ref{Fig-11}. Dashed lines denote the middle of the respective laser pulse, the pulses are shown as shaded area. (bottom) Coulomb momentum transfer vs the simulation time in the laser pulses of a Gaussian form. The final $\langle p_z^C\rangle$ are given with corresponding colors. The full widths at half maximum of the pulse (FWHM) are indicated in the inset. The time-dependent CMT $\langle p_z^C\rangle (t)$ reaches its absolute value maximum at the moment when the electron traverse from the upper half of the $x-z$ plane to the bottom one. }}\label{Fig-12}
\end{center}
\end{figure}
The  pulse shape impacts the resulting effect, which we have analyzed  on the example of  Gaussian pulses.
Before discussing the results for the Gaussian pulse case, let us formulate the key physical points which are advocated in the study with trapezoidal laser pulses. The key findings are that (1) the ionization yield in this regime exhibits oscillating time-dependent  behavior with respect to the pulse duration; (2) This effect is due to the  similar temporal behavior of the CMT and the expectation value of the electron wave packet trajectory;  (3) the oscillating trajectory emerges  due to the competition between the Coulomb attraction and the nondipole drift.

The ionization yield as a function of the laser pulse duration for Gaussian pulses is shown in Fig.~\ref{Fig-11}. While the explicit functional dependence of the yield on the pulse duration deviates from that of trapezoidal pulses, the underlying qualitative physics remains the same. The ionization yield again exhibits a temporal dependence featuring an oscillation cycle,  characterized by a typical time parameter at which the ionization yield is maximized for the given $a_0$, which is the analogue of the half-period of the ionization yield in the trapezoidal case. The similarity of the physics behind is confirmed by the CMT  dependence on the pulse duration, see Fig.~\ref{Fig-11}, which as in the trapezoidal case, is correlated very well with the behavior of the yield: the maximum of the absolute value of the final CMT corresponds to the maximum of the ionization yield.

As in the trapezoidal case, the laser-pulse-duration-dependent CMT is related to the trajectory of the wave packet, which in the case of Gaussian pulse is illustrated in Fig.~\ref{Fig-12}. The oscillation of CMT is correlated with the oscillating behavior of the trajectory [Fig.~\ref{Fig-12} (top)].
The final CMT is built up during the interaction, with the time-dependent CMT characterized by its maximum  and  asymmetry [Fig.~\ref{Fig-12} (bottom)]. These parameters are correlated with the oscillating features of the trajectory, namely, with the sign of $\langle z \rangle$ and the asymmetry of the trajectory, which arise from the competition between the Coulomb and Lorentz forces. This is similar to the origin of the trajectory oscillation in the trapezoidal pulse case.

\begin{figure}
\begin{center}
        \includegraphics[width=0.5\textwidth]{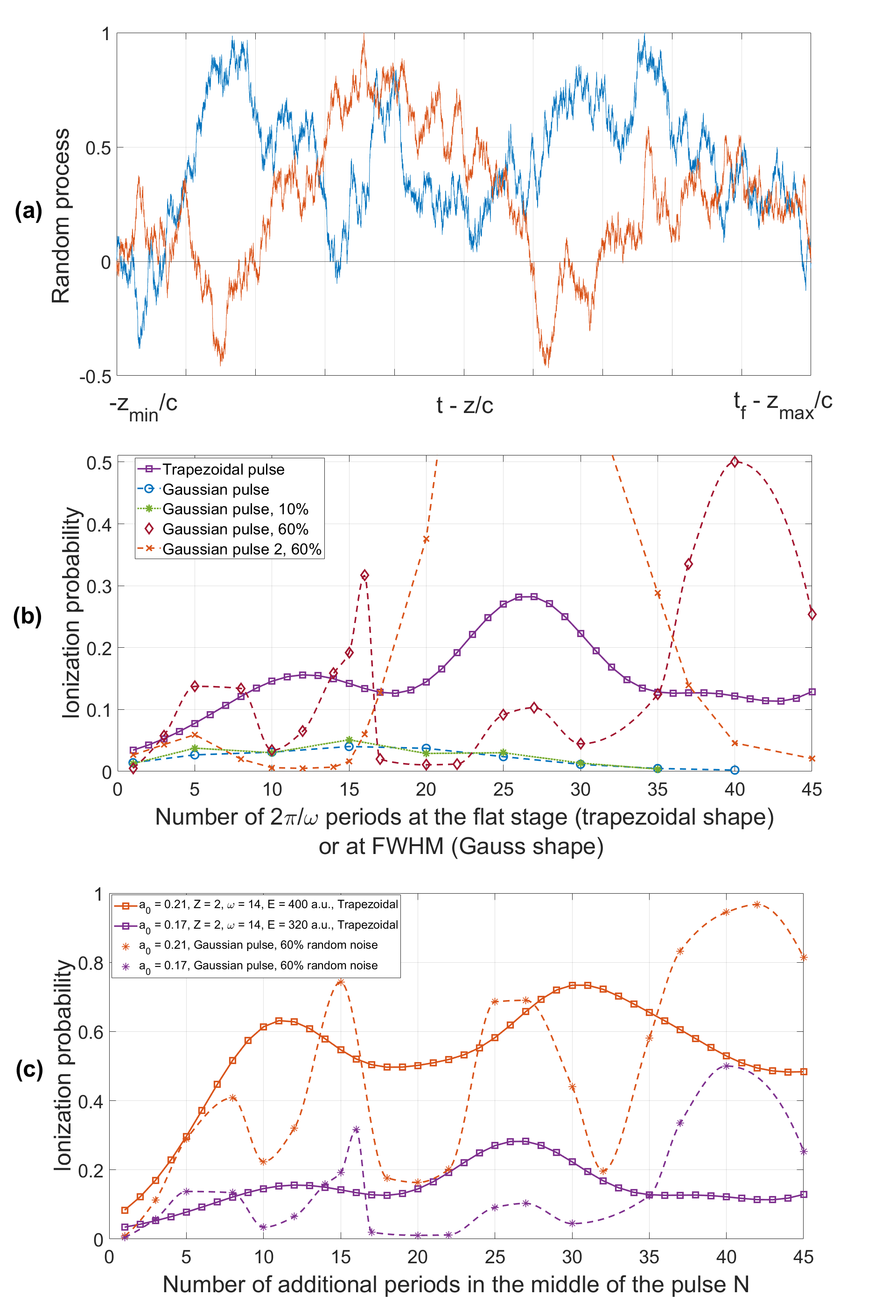}
        \caption{ {  (a) An example of random processes $R$, used to calculate Gaussian pulses with random distortion Eq.~(\ref{random}). The ionization probability vs the laser pulse duration in the nondipole regime for $Z=2$, $\omega =  14 ~\text{a.u.}$: (b)  for $E = 320$ a.u.,  the forms of the laser pulse are trapezoidal, Gaussian (see indications in inset), and randomly distorted Gaussian (with  two forms as in top panel). (c) The cases of $E = 320$ a.u. vs $E = 400$ a.u. where the noise is 60\%.}}\label{Fig-13}
         \end{center}
\end{figure}

Moreover, not only the physical mechanisms underlying the considered effect are consistent across both trapezoidal and Gaussian pulses. When random noise is superimposed on a Gaussian pulse, periodic oscillations in the ionization yield emerge as a function of pulse duration also in this case, mirroring the behavior observed in the trapezoidal case.

The XFEL pulses are known to have random fluctuations. We have analyzed the  influence of the pulse-shape fluctuations, performing new simulations with a randomly distorted Gaussian pulse:
\begin{equation}
\label{random}
    A_r = \left[1 + r_0R(t-z/c) \right] \cdot \sin(\omega (t - z/c)) \cdot e^{\frac{(t - z/c)^2}{2\tau^2}},
\end{equation}
where $r_0$ is a coefficient which we take from $0.1$ to $0.6$, and $R(z-t/c)$ is a normalized random process such that $R(t=t_{\text{start}}) = 0; ~~ R(t=t_{\text{finish}}) = 0; ~~ \max\{ R\} = 1$.
An example of process R is shown in Fig.~\ref{Fig-13} (first row). The ionization probabilities calculated using the pulse defined in Eq.~(\ref{random}) are presented in Fig.~\ref{Fig-13} (second row). For a small parameter $r_0 \sim 0.1$, the probabilities fluctuate around the results obtained with the undisturbed Gaussian pulse. As $r_0$ increases toward  $r_0 \sim 0.6$, the curves in Fig.~\ref{Fig-13} (third row) show the yield oscillation qualitatively similar to that of  the trapezoidal pulse, with two large peaks within the interval $N=(5,30)$. The similar picture is observed  for higher laser intensity in Fig.~\ref{Fig-13}~(third row).

 While at first sight surprising, this behavior has a straightforward intuitive explanation: the primary difference in electron dynamics between a Gaussian and a trapezoidal pulse is the evolution of the drift velocity. In the Gaussian pulse, the drift velocity in the propagation direction builds up gradually throughout the pulse duration, whereas in the trapezoidal case, it remains constant during the plateau part of the pulse. The fluctuations imposed on the Gaussian pulse tend to distort randomly the gradually increasing (decreasing) drift motion, effectively bringing the situation to the trapezoidal case. The calculations using trapezoidal pulses are valuable, as they provide understanding of the main underlying physics.
 }

 \begin{figure*}
    \centering\includegraphics[width=0.65\textwidth]{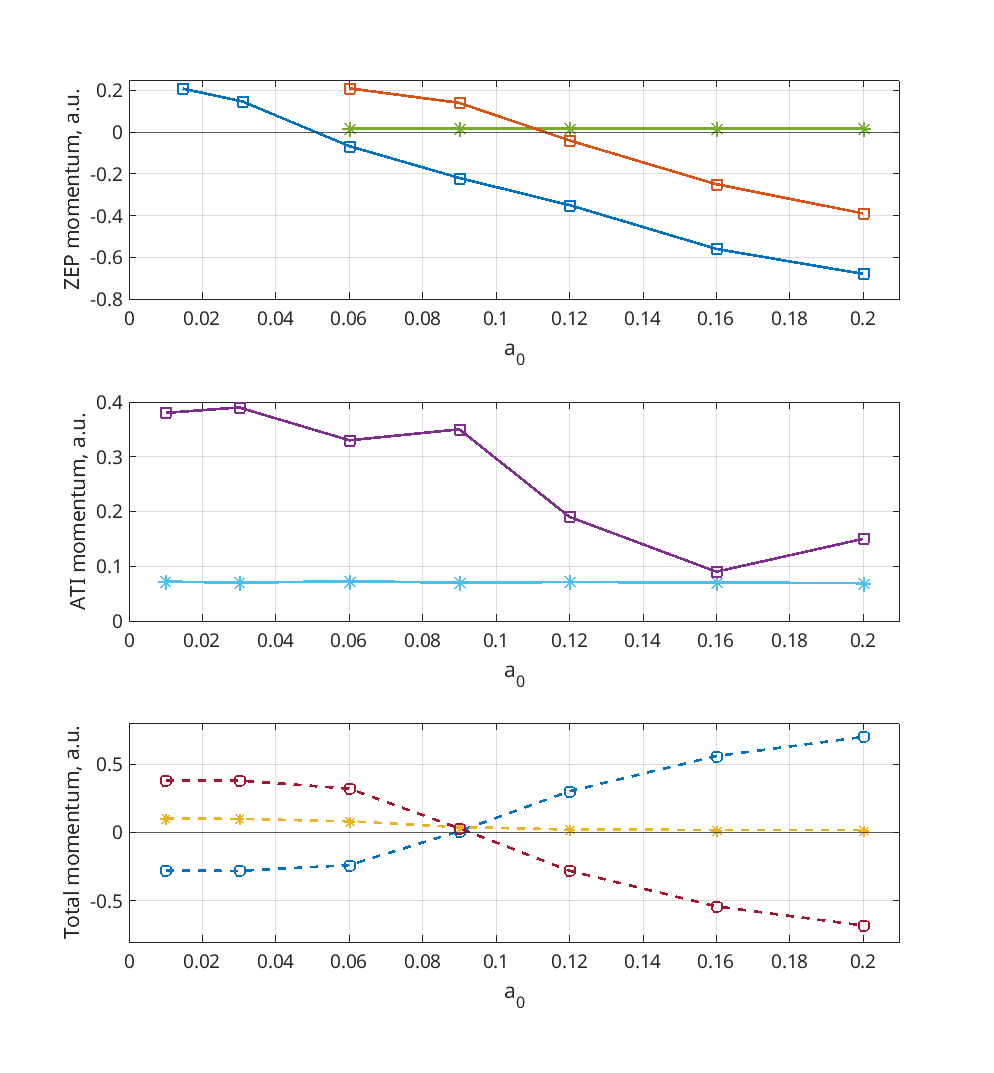}
    \caption{ \textbf{Momentum sharing between the photoelectron and ion vs the field parameter $a_0$.} For the ZEP (first row): the average photoelectron momentum $\langle p_{ze} \rangle$ -- (blue solid line, squares), the total momentum of absorbed photons $\langle n\omega/c\rangle=\langle\varepsilon_e + I_p\rangle/c$ -- (green solid, stars), {the CMT $p_C$, calculated via the pseudo-classical simulation (red solid line, squares).}
       For ATI peaks (second row): $\langle p_{ze} \rangle$ -- (magenta solid, squares); $\langle n\omega\rangle/c$ -- (light blue solid, stars).
               For the total PMD (third row):  $\langle p_{ze} \rangle$ -- (dashed brown line, $\langle n\omega\rangle/c$ -- (dashed yellow line,  circles),   circles), $\langle p_{zi} \rangle$ -- (dashed blue line, squares).  }     \label{fig:pzn}
\end{figure*}

\section{Photon momentum sharing between the photoelectron and ion}\label{sec:momentumSharing}

Here, we address the question of the partition of the momentum of absorbed photons  between  the ionized electron and the emerged ion in the case of the nondipole stabilization regime.

We begin with the energy and momentum conservation, which are fulfilled during the ionization process:
\begin{eqnarray}
     n\omega -I_p&=& \varepsilon_e  + \varepsilon_i \\
         \frac{n\omega}{c} &= & p_{ze} + p_{zi},
\end{eqnarray}
where $\varepsilon_e$ and $\varepsilon_i$, $p_{ze}$ and $p_{zi} $ are the photoelectron and ion kinetic energies and momenta, respectively.  The kinetic energy of the ion can be neglected because of its large mass with respect to the electron, and the photoelectron energy determines the total number of absorbed photons. Then, the ion final momentum is
\begin{equation}
\label{eq:pzn}
    p_{zi} = (\varepsilon_e + I_p)/c - p_{ze}.
\end{equation}

The photoelectron average momentum $\langle p_{ze}\rangle$ and average energy $\langle\varepsilon_e\rangle$ can be extracted directly from our simulation results, with $I_p=Z^2/2$, which are shown in Fig.~\ref{fig:pzn} in the case of a short Gaussian pulse  and different field intensities.  A particular appearance of Fig.~\ref{fig:pzn} depends on the length of the pulse $N$ and its frequency $\omega$, although qualitatively the picture remains similar. We distinguish the momentum sharing for the near ZEP, from that of the ATI multiphoton peaks (mostly a single photon ionization peak is observed), because the physical origin of their momentum partition is different.

For the ZEP, the total momentum of the absorbed photons is approximately vanishing as expected, because it is produced with multiphoton absorption and emission of an equal number of photons. The average photoelectron momentum at large $a_0$ is negative, which is known to emerge due to the CMT during the interaction and producing the angular peak in the laser counterpropagation direction \cite{Foerre_2006}. For the ZEP CMT competes with the laser induced drift, because of which the sign of $\langle p_{ze}\rangle$ changes at low $a_0$. It turns out, that for $a_0 < 0.05$ the CMT is not strong enough to reverse $p_{ze}$. However, as $a_0$ grows, $p_{ze}$ becomes negative, increasing by module at least up to $a_0 \approx 0.25$.

For the ATI peaks, the total momentum of the absorbed photons is approximately $  \omega /c\approx 0.1$ as expected. The average photoelectron momentum is positive and decreasing with larger $a_0$, $\langle p_{ze} \rangle >  \omega/c $, which means that the ion momentum is negative (opposite to the photon momentum) and also decreasing with larger $a_0$.

We calculated also the $\langle p_{ze}\rangle$ via the total PMD. It is known that for the low $a_0$, the ionization is  dominated by the ATI process \cite{Telnov_Chu_2021}, and consequently, $\langle p_{ze}\rangle$ via the total PMD coincides with that of the ATI.
However, for $a_0 > 0.1$ the Coulomb scattering takes over, the peak dominates, which results in a change of sign for $p_{ze}$. At some point $a_0 \approx 0.1$, ATI contribution and the ZEP contribution become equal, which results in a nearly zero photo-electron momentum along the laser propagation direction.

Finally, in order to prove that $p_{ze} < 0$ for the ZEP  is caused by  CMT, we carried out a simple classical simulation. We create a classical ensemble of electrons mimicking the final electron wave packet. Both final coordinates $\{x_f,z_f\}$ and $\{p_{x_f},p_{z_f}\}$ are sampled from the ZEP wave function density $\psi(x,z,t_f)$.  $\{p_{x_f},p_{z_f}\} = -i \nabla \psi(x_f)/\psi(x_f)$. Each trajectory is propagated backward in time  classically until its energy $\varepsilon (t) = [p(t) - p_L(t)]^2/(2m) + V < 0$, where $p_L(t)$ is a momentum of the same electron backward propagated only with the laser field. That point $t_i$ is considered a moment of ionization. The CMT is calculated as $p_C = - \int_{t_i}^{t_f} \nabla V(x(t'),z(t'))dt'$. The CMT average over the ensemble is shown {  in first row} in Fig.~\ref{fig:pzn}. The CMT with the opposite sign represent the momentum transferred to the ion. As the total photon momentum for ZRP is vanishing, one expects the CMT to coincide with $p_{ze}$. While our simple estimate does not give coincidence, nevertheless it reproduces the trend of an actual $p_{ze}$, deviating  only by a constant factor.

\section{Conclusion}

We have developed a method for the numerical solution of the Foldy-Wouthuysen transformed TDDE  for the aim of treating the strong field ionization in the relativistic regimes. Our method is based on the coordinates scaling ansatz that absorbs as well the kinetic propagation phase of the wave function. The method is tested and analyzed in Ref.~\cite{Boitsov_2025}. Using this method, we solve
numerically the 2D problem of the strong field ionization of a hydrogen-like helium atom in
a high-frequency strong laser field. We have found a previously unobserved effect that the ionization
yield in the stabilization regime oscillates with the laser pulse duration. While
this effect exists in the dipole regime as well as in the nondipole one the underlying physical mechanism  and the features of the yield oscillation  are  different.

In the dipole case, the yield oscillation has been explained  by means of the dynamic interference phenomenon, combined with the electron periodic dynamics in the polarization direction in the KH potential.

In the nondipole case the ionization yield oscillation is a result of a more complex dynamics. The key point is the periodic slow oscillation along the laser propagation direction, which arises due to the competition between the laser induced drift and the Coulomb attraction. This slow oscillation brings about the varying average energy after the interaction and varying CMT, which finally results in the periodic variation of the ionization yield  depending on the interaction time. We have estimated the scaling of the oscillation period and the threshold intensity when the drift dominates the ionization.

While in the nondipole regime,  still  the dynamic interference existed, however with the diminished role, the significant part of ionization emerges due to the asymmetric Coulomb momentum transfer to the electron wave packet during the continuum dynamics dominated by the nondipole drift.   At first sight it seems the larger the nondipole drift, the  larger is the asymmetry, and the larger should be the CMT. However, this picture overlooks the long term role of the Coulomb field for the electron dynamics in the continuum. Owing to the Coulomb field the electron wave packet  features  slow oscillations along with the nondipole drift. During turning points of this Coulomb orbiting the electron slows down and stay a long time close to the core, acquiring a large CMT, which results in increasing the anomalous lobe in PMD and in the total ionization yield. This happens periodically when the duration of the laser pulse matches the multiples of the Coulomb orbiting period.

{  The considered effect of the  ionization yield oscillation is quantitatively dependent on the laser pulse shape. It is more regular in trapezoidal laser pulses, and has irregularities in the randomly distorted Gaussian pulses. However, the underlying physical origin remains the same.  }

We have investigated also the absorbed photon momentum partition between the photoelectron and ion during the ionization in the nondipole stabilization regime. There are two characteristic features to underline. Firstly, the momentum partition is very different for the ZEP and ATI peaks. Secondly, the large Coulomb effect results in the ZEP photoelectron final average momentum being opposite to the laser propagation direction at large $a_0$, which however is reverted when $a_0\ll 1$.

{  With a hydrogen target the considered effect can be observed with a photon energy of about 50 eV (wavelength about 20 nm). This kind of photon beam is available in e.g. the DESY FLASH facility. A typical energy of the beam is about $\sim 500$ $\mu$J and the beam duration $\sim 10$ fs which amounts in a average power of $5\times 10^{10}$ W. In the case of the beam radius of 0.3 $\mu$m, the intensity will reach $5\times 10^{19}$~W/cm$^2$ ($a_0\approx 0.17$) when the considered effect will take place. While the standard focal spot from the FLASH beam is 10 -- 50 $\mu$m, extreme focusing below 1~$\mu$m	 is still possible with the specialized nanofocusing optics, see Refs.~\cite{LCLS,flash,Dziarzhytski_2016,Vassholz_2021,Tong_2022}. Thus, the parameters used in this paper are very demanding to realize in the present-day XFEL facilities, but they are  not prohibitive. The conditions for the measurement of the given effect are similar to those which are required to observe the counterintuitive lobe in the angular distribution of Ref.~\cite{Forre_2006}.
The signature of the slow Coulomb orbiting during the continuum dynamics can be observed in the total ionization yield measured in the laser pulses of different duration, even without precision measurement of the photoelectron momentum distribution. Especially interesting will be the measurement of the nontrivial momentum partition of the absorbed photons between the photoelectron and the resulting ion. }

\appendix

\section{Calculation methods} \label{sec:methods}

We consider a hydrogen-like ion in a strong laser field, describing the electron dynamics with the Dirac equation:
\begin{eqnarray}
\label{eq:Dirac}
    i\partial_t \psi &=& H_D \psi \\
    H_D&=& \beta c^2 + O + V,\nonumber\\
    O&=&  c  \boldsymbol{\alpha}  \left( \mathbf{p} -\frac{e}{c} \mathbf{A}(\mathbf{r},t) \right)\nonumber\\
    \textbf{A}(\mathbf{r},t )&=& A(t-z/c) \textbf{e}_x ,\nonumber
\end{eqnarray}
where $\boldsymbol{\alpha},\,\beta$ are the Dirac matrices, $\mathbf{A}(\mathbf{r},t)$ is the vector potential of the laser field,   $V(\mathbf{r})=-Z/r$ is the atomic potential, with the charge $Z$, and $e$ is the electron charge.
We account for the nondipole effects using the laser vector potential as a propagating wave.
To employ the coordinate scaling method  \cite{Boitsov_2025}, we apply the FW transformation to the Dirac Hamiltonian in Silenko's (quasiclassical) form \cite{Silenko_2008} as an $\hbar$ expansion:
\begin{equation}
    \psi_{FW} = U_{FW}\psi = e^{iS_{FW}}\psi,
\end{equation}
arriving at the FW Hamiltonian \cite{Boitsov_2025}:
\begin{equation}
\label{eq:HFW}
    H_{FW} = \beta \varepsilon + V - \frac{1}{8}\left\{ \frac{1}{\varepsilon(\varepsilon + c^2)}, [O,[O,\mathcal{F}]] \right\} + \mathcal{O}(\lambdabar^2)
\end{equation}
where $\lambdabar$ is the Compton wavelength, and
\begin{equation}
\label{eq:HFW_O}
    \begin{split}
    \mathcal{F} = V - i\hbar & \partial_t, ~~~    \varepsilon = c^2 \sqrt{1+\frac{O^2}{c^4}}.\nonumber
    \end{split}
\end{equation}
Introducing the magnetic field $\mathbf{B} = \mathbf{\nabla\times A} = \partial_zA(t,z) \hat{\textbf{e}}_y$, the operator $\varepsilon$  reads
\begin{equation}
\label{eq:epsilon2D}
    \varepsilon = c^2 \sqrt{1 + \frac{\left(p_x-\frac{e}{c}A(t,z)\right)^2}{c^2} + \frac{p_z^2}{c^2} - \frac{e\hbar}{c^3}\Sigma_y \cdot B_y}
\end{equation}
where $\Sigma$ is the spin operator. Taking into account $\partial_zA(t,z) \sim \omega A(t,z)/c$, and the smallness of the parameter $a_0\lambdabar/\lambda$ for the applied parameters, with the laser wavelength $\lambda$, we obtain the expansion:
\begin{eqnarray}
    \varepsilon &=& c^2 \varepsilon_0(\textbf{p},t) - \frac{\lambdabar}{2\varepsilon_0(\textbf{p},t)} \Sigma_y \cdot B_y + \mathcal{O}\left(\left( a_0 \lambdabar/\lambda\right)^2\right),\\
  &&  \varepsilon_0(\textbf{p},t)= \sqrt{1 + \frac{\left(p_x-\frac{e}{c}A(t,z)\right)^2}{c^2} + \frac{p_z^2}{c^2}}.\nonumber
\end{eqnarray}
 As a result, there are no more matrices inside the square-root operator $\varepsilon_0$, which simplifies the calculations. However, the variables from different coordinate spaces, namely momenta $p_x$ and $p_z$ are still mixed with the regular coordinate $z$.  In order to  implement the square-root operator, we  expand it into a Taylor series:
 \begin{eqnarray}
\label{eq:TaylorRoot}
    \varepsilon_0(\textbf{p},z,t)  &= &\sqrt{1 + X(\textbf{p},z,t) } \nonumber\\
    &= &1 + \frac{1}{2}X(\textbf{p},z,t) - \frac{1}{8} X(\textbf{p},z,t)X(\textbf{p},z,t)  + \dots .
\end{eqnarray}
Note, that the term $X^2(\textbf{p},t)$ produces terms, proportional to $\partial_zA(t,z)$ and $\partial^2_{zz}A(t,z)$ and so on. Thus, we single out the terms by the different orders of $\partial_z^nA(t,z)$ and then roll back the expanded expressions  inverting Eq.~\eqref{eq:TaylorRoot}):
\begin{widetext}
\begin{eqnarray}
\label{eq:Epsilon0}
     \varepsilon_0
     &= &\sqrt{1 + \frac{\left(-i\partial_x-\frac{e}{c}A(t,\tilde{z})\right)^2}{c^2} - \frac{\partial_{zz}^2}{c^2}}  -\frac{i}{2}\frac{1}{c^2}\frac{e}{c}\frac{\partial A(t,z)}{\partial z}  \left[\frac{\left(-i\partial_x-\frac{e}{c}A(t,\tilde{z})\right)}{c} \frac{(-i \hbar \partial_z)}{c}\right] {\left[1 + \frac{\left(-i\partial_x-\frac{e}{c}A(t,\tilde{z})\right)^2}{c^2} - \frac{\partial_{zz}^2}{c^2}\right]^{-3/2}} + \mathcal{O}\left(\left( a_0 \lambdabar/\lambda\right)^2\right),\nonumber\\
       \end{eqnarray}
where the designation $z=\tilde{z}$ means that derivative operators do not act on $A(t,z)$. In Eq.~\eqref{eq:Epsilon0} we omit terms, proportional to $(\partial_zA(t,z))^2$ and  $\partial^2_{zz}A(t,z)$ as of a high-order smallness. For the implementation of the first term in  Eq.~\eqref{eq:Epsilon0}, we again apply the Taylor expansion method and
obtain an expression similar to the 1D case  of Ref.~\cite{Boitsov_2025}:
\begin{eqnarray}
\label{eq:epsilonTerm2D}
     && \sqrt{1 + \frac{\left(-i\partial_x-\frac{e}{c}A(t,\tilde{z})\right)^2}{c^2} - \frac{\partial_{zz}^2} {c^2}} ~ \phi(x,z)~e^{i\varphi(x,z)} \\
     &=&\phi T_0(x,z,t)
    -  i  \lambdabar \biggl[\phi'_x   T_{1,x}(x,z,t) + \phi'_z T_{1,z}(x,z,t)\biggr]\nonumber
    -\frac{ \lambdabar^2}{2} \biggl[ \phi''_{xx} T_{2,xx}(x,z,t) + 2 \phi''_{xz} T_{2,xz}(x,z,t) + \phi''_{zz} T_{2,zz}(x,z,t) \biggr]
    + \mathcal{O} \left( \lambdabar^3 / \ell^3 \right),\nonumber
\end{eqnarray}
where we assume that $\phi' \sim 1/\ell$ and $\lambdabar / \ell \ll 1$, i.e. function $\phi$ does not contain large relativistic oscillations. The second term can be written in the same manner
\begin{equation}
\label{eq:epsilon2Term2D}
\begin{split}
     & \left[\frac{\left(-i\hbar\partial_x-\frac{e}{c}A(t,\tilde{z})\right)}{mc} \frac{(-i \hbar \partial_z)}{mc}\right] {\left[1 + \frac{\left(-i\hbar\partial_x-\frac{e}{c}A(t,\tilde{z})\right)^2}{m^2c^2} - \frac{\hbar^2\partial_{zz}^2}{m^2c^2}\right]^{-3/2}}
          =\phi W_0(x,z,t)
    -i  \lambdabar \biggl[\phi'_x   W_{1,x}(x,z,t) + \phi'_z W_{1,z}(x,z,t)\biggr] + \mathcal{O} \left( \lambdabar^2 / \ell^2 \right)
\end{split}
\end{equation}
where we omit some $\lambdabar^2$ terms which have an additional smallness via $\omega/c^2$.

Recalling Eq.~\eqref{eq:HFW} and using Eqs.~\eqref{eq:epsilon2D} and \eqref{eq:Epsilon0},  the FW Hamiltonian reads:
 \begin{eqnarray}
\label{eq:HFW2D}
    H_{FW} &=& \beta c^2 \varepsilon_0  -   \frac{e\lambdabar}{2\varepsilon_0} \beta \Sigma_y \cdot B_y+ V +\frac{e\lambdabar}{8c^3} \left\{ \frac{1}{\varepsilon_0(\varepsilon_0 + 1)}, \left[ \boldsymbol{\Sigma}\cdot(\boldsymbol{\pi} \times \tilde{\mathbf{E}} - \tilde{\mathbf{E}} \times \boldsymbol{\pi}) -  \Delta V \right]\right\} + \mathcal{O}(\lambdabar^2/\ell^2)
\end{eqnarray}
where $\boldsymbol{\pi} = \boldsymbol{p} - \frac{e}{c}\mathbf{A}$, and $\tilde{\mathbf{E}} = \nabla V - \frac{1}{c}\dot{\mathbf{A}}$.
With the Hamiltonian Eq.~\eqref{eq:HFW2D} and factorization $\psi = \phi e^{i\varphi}$, we  arrive to the FW governing equation
\begin{equation}
    \label{eq:muFWEqScaled}
    \begin{split}
    i  \partial_t \phi =   c^2 \biggl[ {T}_0(\xi R,t)&
    - i  \lambdabar \frac{\boldsymbol{T}_1}{R} \cdot \boldsymbol{\partial_{\xi}}
    - \frac{ \lambdabar^2}{2 } \frac{\boldsymbol{T}_2}{R^2} \boldsymbol{\partial_{\xi\xi}}^2 + \mathcal{O} \left(\frac{\lambdabar^3}{\lambdabar_e^3}\right) \biggr]
    \phi
    + \frac{i}{2}\frac{\hbar e}{mc} \frac{\partial A(t,z)}{\partial z}  \biggl( 1 + i\beta \Sigma_y \biggr)\biggl[ W_0(x,z,t)
    -i  \lambdabar \boldsymbol{W}_1 \cdot \boldsymbol{\partial_{\xi}} +\mathcal{O} \left(\frac{\lambdabar^2}{\lambdabar_e^2}\right) \biggr] \phi + \\
    + \frac{\lambdabar e}{8 c^3} \biggl\{ & \frac{1}{\varepsilon_0^2(x,z,t) +\varepsilon_0(x,z,t)}, \left[ \boldsymbol{\Sigma}\cdot(\boldsymbol{\pi} \times \tilde{\mathbf{E}} - \tilde{\mathbf{E}} \times \boldsymbol{\pi}) -  \Delta V \right]\biggr\} \phi
    + V(\xi R)\phi +  \phi \left[\dot{\varphi} - \frac{\dot{R}}{R}\boldsymbol{\xi} \cdot \boldsymbol{\partial_{\xi}} \varphi\right] + i\frac{\dot{R}}{R}\boldsymbol{\xi} \cdot \boldsymbol{\partial_{\xi}} \phi
    + \mathcal{O} \left( \lambdabar^2\right)
    \end{split}
\end{equation}
where $\phi(x,z,t)$ is a smooth envelope function, $\phi' \sim 1/\ell$, $T_i$ and $W_i$ are smooth functions given in Appendix B in \cite{Boitsov_2025}.
\end{widetext}
Further, we  introduce the smooth  phase in momentum space $\varphi(t,x,z)$ ( $\varphi \gg \varphi'(t,x,z) \lambdabar$), using the following  educated guess:
\begin{equation}
\label{eq:phiP2D}
    \hat{\varphi}(p_x,p_z,\tilde{z},t) = - \int^t_{0} \sqrt{1 + \frac{\left(p_x-\frac{e}{c}A(t',\tilde{z})\right)^2}{m^2c^2} + \frac{p_{z}^2} {m^2c^2}}dt'.
\end{equation}
Since  $ \omega \ll c^2$ and $\hat{\varphi}$ has to be smooth,  it is enough to calculate Eq.~\eqref{eq:phiP2D} for a few reference values $\tilde{z}$ and then interpolate to obtain $\hat{\varphi}(p_x,p_z,z,t)$ for any $z$. Interpolation can be done via, for example, splines, or with Chebyshev polynomials, because we are free to choose reference nodes $\tilde{z}$.

\begin{figure}
       \includegraphics[width=0.5\textwidth]{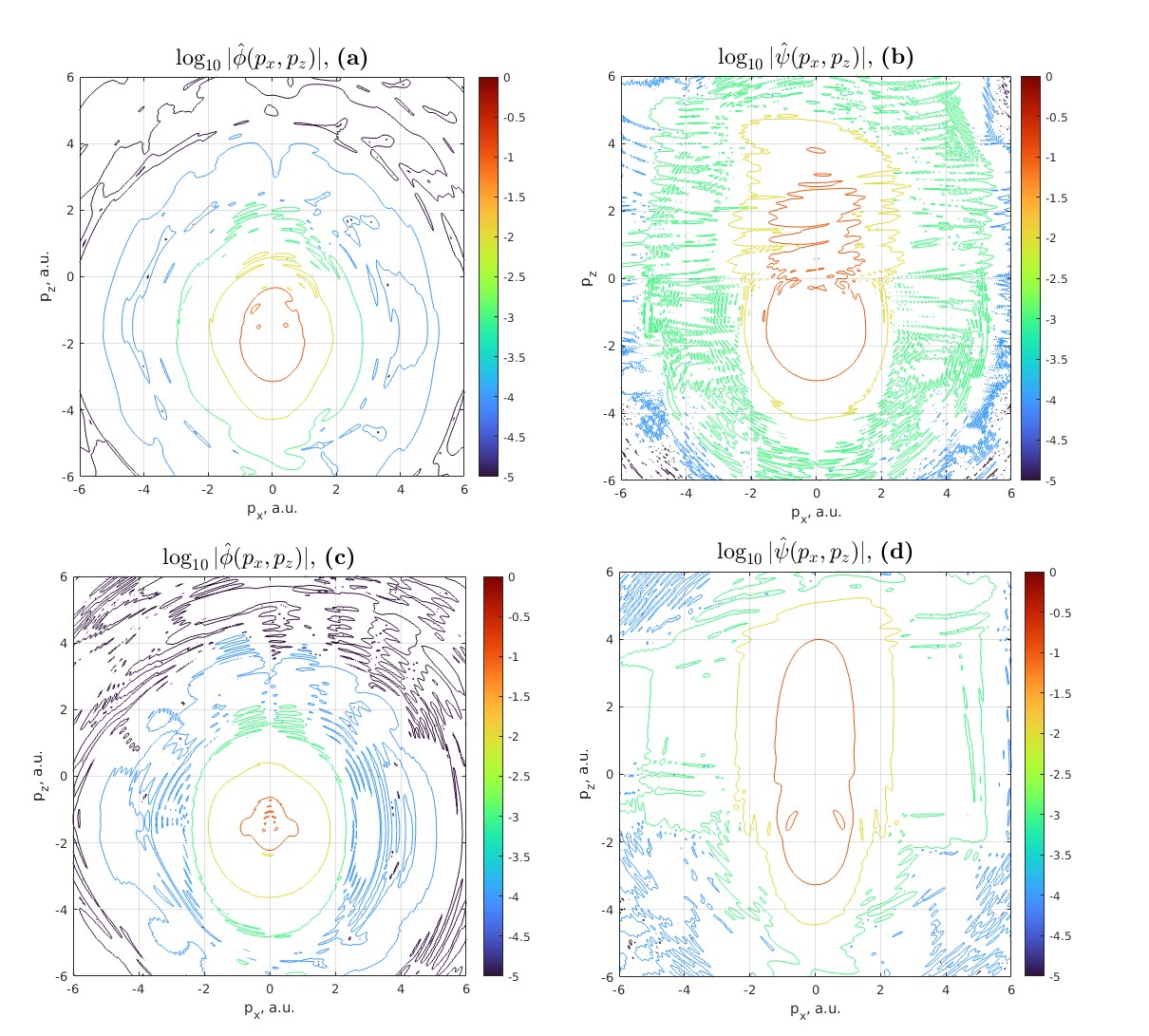}
    \caption{ The comparison of the envelope function $|\hat{\phi}(\mathbf{p})|$ and the original wavefunction  $|\hat{\psi}(\mathbf{p})|$. Parameters of the field: $\omega = 14$~a.u., $E = 400$~a.u., $a_0 = 0.21$. (a) $\hat{\phi}(\mathbf{p})$ after 8 cycles, (b) $\hat{\psi}(\mathbf{p})$ after 8 cycles, (c) $\hat{\phi}(\mathbf{p})$ after 30 cycles, (d) $\hat{\psi}(\mathbf{p})$ after 30 cycles}
    \label{figTheor1}
\end{figure}

As a result, the procedure for the phase $\varphi$ calculation at each time step $\Delta t$ is the following:
\begin{enumerate}
    \item Calculate \\$\hat{\varphi}(p_x,p_z,\tilde{z},t) = \int_{0}^{t} \sqrt{1 + \frac{\left(p_x-\frac{e}{c}A(t',\tilde{z})\right)^2}{m^2c^2} + \frac{p_{z}^2} {m^2c^2}}dt'$ for every reference value $\tilde{z} \in [-L_z, L_z]$.
\item Interpolate $\hat{\varphi}(p_x,p_z,\tilde{z},t)$ to get $\hat{\varphi}(p_x,p_z,z,t)$.
\item Calculate critical points $x_0(p_x,p_z) = - \frac{\partial\hat{\varphi}}{\partial p_x}$ and $z_0(p_x,p_z) = - \frac{\partial\hat{\varphi}}{\partial p_z}$.
\item Calculate the phase as $\varphi(x)|_{x=x_0(p)} = \hat{\varphi}(p) { p x_0(p)}$.
\item Rescale $x' = \beta(t)x$  and $z' = \beta(t)z$. Since the set of points $\{x_0,z_0\}$ is non uniform and does not make up a grid but rather a mesh, it complicates an interpolation $\varphi(x,z)\rightarrow \varphi(x',z')$. One can use different technics for the solving of optimization problem in order to interpolate from the 2D non uniform mesh to the other non unifrom mesh.
\end{enumerate}
This algorithm  provides us with a decent choice for the smooth phase $\varphi$. In Fig.~\ref{figTheor1} the module of the smooth envelope function $\hat{\phi}(\mathbf{p})$ and the wave function $\hat{\psi}(\mathbf{p})$ are presented. One sees that  $\hat{\phi}(\mathbf{p})$ is more narrow than $\hat{\psi}(\mathbf{p})$, and located closer to the origin, which means that in the coordinate space function $\phi(\mathbf{x})$ is much more smooth than the wavefunction $\psi(\mathbf{x})$.

We have carried out relativistic simulations via Eq.~\eqref{eq:muFWEqScaled} and compare those with the results of the nondipole TDSE with the  Hamiltonian:
\begin{equation}\label{eq:TDSEnon-dip}
    H_{\text{nd}} = \frac{1}{2}\biggl[p_x - \frac{e}{c}A(t,z)\biggr]^2 + \frac{1}{2}p_z^2 + V(r).
\end{equation}
{The linearly polarized laser pulse is of a trapezoidal form, with the  vector potential  $A_x(\eta)=A_0 \sin(\omega \eta)$ in the flat part of the pulse, and with the rising and falling edges described as $A_x(\eta)  = A_0 e^{  -2   \ln(2)   \left( \eta/\tau\right)  } \sin(\omega \eta)$, where $\eta = t - z/c$ for the nondipole and $\eta = t$ for the dipole cases.}

\bibliography{strong_fields_bibliography}

\end{document}